\documentclass[acmsmall]{acmart}

\usepackage{flushend}
\usepackage{float}
\usepackage{subfigure}
\AtBeginDocument{%
  \providecommand\BibTeX{{%
    \normalfont B\kern-0.5em{\scshape i\kern-0.25em b}\kern-0.8em\TeX}}}
\usepackage{tabularx}
\usepackage{bbding}
\usepackage{graphicx}
\usepackage{caption}
\usepackage{geometry}
\usepackage{amsmath}
\usepackage{makecell}
\usepackage{float}
\usepackage{color}
\usepackage{booktabs}
\usepackage[normalem]{ulem}
\usepackage{tabu}
\usepackage{hyperref}
\usepackage{xcolor}
\usepackage{appendix}

\usepackage{graphicx}  
\usepackage{booktabs}  

\newcommand{\added}[1]{{#1}}
\newcommand{\removed}[1]{}

\author{Long Ling}
\authornote{These authors contributed equally to this paper.}
\email{lucyling0224@gmail.com}
\orcid{0009-0001-2635-788X}
\affiliation{
  \institution{Intelligent Big Data Visualization Lab, Tongji University}
  \city{Shanghai}
  \country{China}
}

\author{Xinyi Chen}
\authornotemark[1] 
\email{chenxinyi810musi@gmail.com}
\orcid{0009-0006-0121-5355}
\affiliation{
  \institution{Zhejiang University}
  \city{Hangzhou}
  \country{China}
}

\author{Ruoyu Wen}
\authornotemark[1]
\email{rwe77@uclive.ac.nz}
\orcid{0009-0008-0052-0045}
\affiliation{
  \institution{University of Canterbury}
  \city{Christchurch}
  \country{New Zealand}
}

\author{Toby Jia-Jun Li}
\email{toby.j.li@nd.edu}
\orcid{0000-0001-7902-7625}
\affiliation{
  \institution{University of Notre Dame}
  \city{Notre Dame}
  \country{United States}
}

\author{RAY LC}
\authornote{Correspondence should be addressed to LC@raylc.org.}
\email{LC@raylc.org}
\orcid{0000-0001-7310-8790}
\affiliation{
  \institution{School of Creative Media, City University of Hong Kong}
  \city{Hong Kong}
  \country{Hong Kong}
}

\begin{document}

\setcopyright{acmlicensed}
\acmJournal{PACMHCI}
\acmYear{2024} 
\acmVolume{8} 
\acmNumber{CHI PLAY} 
\acmArticle{337} 
\acmMonth{10}
\acmDOI{10.1145/3677102}

\title[Sketchar]{Sketchar: Supporting Character Design and Illustration Prototyping Using Generative AI}

\begin{abstract}

Character design in games involves interdisciplinary collaborations, typically between designers who create the narrative content, and illustrators who realize the design vision. However, traditional workflows face challenges in communication due to the differing backgrounds of illustrators and designers, the latter with limited artistic abilities. To overcome these challenges, we created Sketchar, a Generative AI (GenAI) tool that allows designers to prototype game characters and generate images based on conceptual input, providing visual outcomes that can give immediate feedback and enhance communication with illustrators' next step in the design cycle. We conducted a mixed-method study to evaluate the interaction between game designers and Sketchar. We showed that the reference images generated in co-creating with Sketchar fostered refinement of design details and can be incorporated into real-world workflows. Moreover, designers without artistic backgrounds found the Sketchar workflow to be more expressive and worthwhile. This research demonstrates the potential of GenAI in enhancing interdisciplinary collaboration in the game industry, enabling designers to interact beyond their own limited expertise.

\end{abstract}



\received{February 2024}
\received[revised]{June 2024}
\received[accepted]{July 2024}



\maketitle
\vspace{-0.3cm}
 \begin{figure}[htbp]
  \centering
  \includegraphics[width=0.99\linewidth]{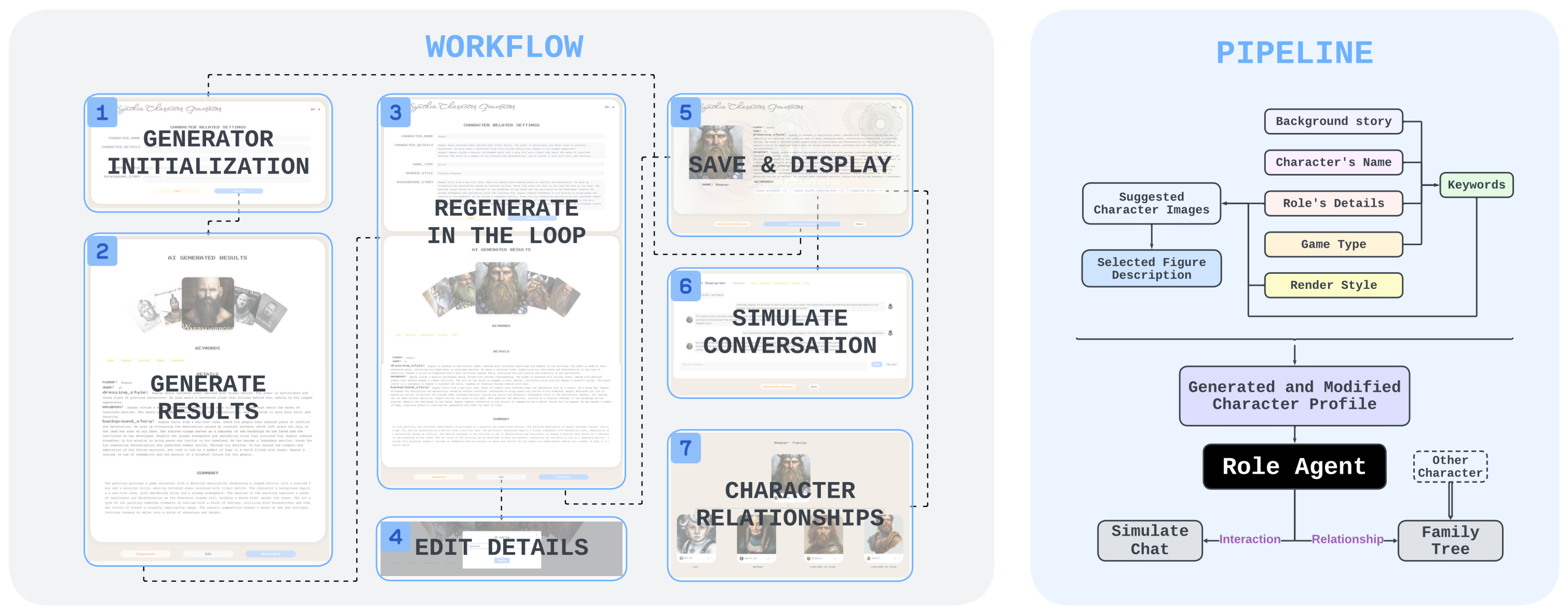}
  \caption{Sketchar is a creative tool for game designers to generate game character profiles using natural language. \textbf{Workflow:} (1) (2) The designer gives initial specifications and obtains generated results, (3) (4) Designer can regenerate and edit the profile, (5) (6) (7) Designers and illustrators can engage with it through actions such as showcasing an ID card, simulating conversations, and editing the relationships between characters. \textbf{Pipeline:} Through the hierarchical content generation with prompt engineering, Sketchar creates character profiles with the help of interactions with LLM.}
  \label{ABS}
    \vspace{-0.36cm}
\end{figure}

\section{Introduction}\label{sec:Introduction}
In game design, the process of collaboratively crafting characters serves as a pivotal point where creative and conceptual objectives intersect~\cite{kuntjara2017character, 9864747, lee2021character}. This process includes the two overlapping stages of character concept development and character artwork illustrations. Thus, game character design is a collaboration between the creative visions of designers and the illustrative skills of artists~\cite{9864747}.

The partnership between designers and artists is full of communication obstacles. Designers express their creative ideas using written explanations, using visual elements only as supplements to the conceptual explication. Afterwards, illustrators convert these inputs into concrete visual prototypes~\cite{9864747}. However, the exchange of ideas between these two domains of creativity may result in misinterpretation, resulting in repetitive cycles of changes that require significant time and resource investment~\cite{9864747}, which is particularly troublesome in the video game industry ~\cite{ko2022we,chung2022artist,chung2023artinter}.

To help game designers overcome communication challenges and lack of expertise in illustration, we implemented a web-based creative tool to allow designers to prototype the character design process both conceptually and visually, easing them into the next step of working with illustrators. Our system takes advantage of the ability of Generative AI (GenAI) tools to give idea inspirations and customized image generation in response to natural language. We also applied ChatGPT to facilitate the refinement of keywords summarized from text input by designers, much like summarizing design documents to make abstract concepts more concrete~\cite{mansimov2015generating}. We input these keywords into the GenAI tool DALLE as prompts to generate reference images for prototyping early-phase character illustrations. The tool developed allows designers to organize and visualize their ideas without expert skills in illustration, creating prototyped character designs that serve as a bridge in communication between designer and illustrator workflows~\cite {angert2023spellburst, verheijden2023collaborative}. The design and testing of the Sketchar tool addresses the following research questions:



\textbf{RQ1:}
What are the challenges to collaboration and communication between designers and illustrators in the team character design process?

\textbf{RQ2:} 
How may we facilitate both narrative and artistic aspects of character design in one workflow by applying GenAI tools for both conceptual and illustrative purpose?

Answering RQ1 allows us to design a tool with data informed from the actual workflows of designers, whereas answering RQ2 allows us to facilitate the designers' ultimate task of communicating with illustrators and artists to implement the design. In this study, we carried out formative interviews with professional game designers, identifying the obstacles in the real-world game design and collaboration workflow within the industry. Based on these findings and insights, we created Sketchar, a tool that uses ChatGPT and DALLE to generate keyword suggestions and visual prototypes, also designed to facilitate conceptual inspiration and character template creation.

We then conducted a mixed-method study with experienced game designers to assess the challenges, constraints, and advantages of working with the tool. We found that the generated images can foster expression of design details, and are particularly satisfactory to those without professional artistic backgrounds. Meanwhile, designers with artistic backgrounds have a higher demand for the quality of the prototypes. Comparing the use of Sketchar against a baseline control of sketching-only, we found that participants reported significantly higher ratings of the collaboration dimension in the Creativity Support Index (CSI). Additionally, we conducted an external evaluation of the output quality of Sketchar from an artistic perspective with professional art designers, showing that they can potentially work as part of real-world workflows.  Our work highlights the way GenAI tools can empower creative processes for those lacking in skill expertise.

\section{Related Work}\label{sec:Background}

\subsection{Character Design in Games}


Character design is a pivotal element in storytelling, influencing the success of games and movies \cite{fogelstrom2013investigation}. In role-play games, immersive experiences rely on meticulously designed characters with unique names, visual representation, and specific features such as personality, gameplay mechanics, logos, and other distinguishing traits \cite{lankoski2004character,canheti2019case}. The process of character creation, often perceived as linear, though actually involves an iterative prototyping approach, progressing through initial concept development, environment creation, and final visualization \cite{fogelstrom2013investigation,voimala2023creating,nop2019development}.

During the creation stages, designers employ visual metaphors to establish a character notion, integrating game-play style and psychological attributes \cite{solarski2012drawing}. Jungian archetypes and narrative content like Joseph Campbell's book 'The Hero with a Thousand Faces' guide the communication of cohesive character concepts \cite{krawczyk2006game,mark2001heroi}. Visual artists contribute during the visual design stage, focusing on visual attributes like body shapes, poses, facial expressions, and color schemes, which are then used to create a playable prototype \cite{lilly2017big,canheti2019case}.

Despite this organized approach, character creation lacks a rigid procedure to be used at every step \cite{lebowitz2011interactive}. In some procedures, iterative user testing is often encouraged and found adaptable to refine characters \cite{martins2021character, tavares2011drawing}. Nevertheless, it should be emphasized that character design extends beyond only visual aspects to align with the narrative and motifs of the game \cite{martins2021character}. Hence, establishing a meaningful connection between a character and the narrative of a game is a significant obstacle in character creation. Our program facilitates the construction of game characters by including input parameters that define game genres, styles, worldviews, and other relevant factors. This allows the created virtual characters to align with the specific context of the game.

\vspace{-0.3cm}

\subsection{Previous GenAI Interaction Studies}

The evolution of Text-to-Image (TTI) machine learning, initiated by the work of Mansimov \cite{mansimov2015generating}, has witnessed significant breakthroughs, from generative adversarial networks (GANS) \cite{zhang2017stackgan,hong2018inferring} to CLIP (Contrastive Language-Image Pretraining) \cite{radford2021learning}. Recent models like DALLE-2 and Stable Diffusion have demonstrated diverse capabilities in generating high-quality pictures, with Stable Diffusion excelling in character face production \cite{borji2023generated,fernandez2022technology}.

Progress in GenAI introduces prompt engineering as a novel interaction paradigm, offering effective human interaction with text-to-image tasks \cite{branwen2020gpt}. Various prompting paradigms, such as the "pre-train, predict, and prompt" for prediction tasks, have emerged \cite{liu2023pre}.  Deckers' approach reframes prompt engineering as interactive text-based retrieval on an "infinite index" \cite{deckers2023infinite}, while Ramesh proposes a similar transformer-based approach for this task \cite{ramesh_zero-shot_2021}. These paradigms have found practical applications, including the use of instant modifiers in "point-in-time engineering" \cite{oppenlaender2022taxonomy}, yet game design remains an underserved area.

Recent technical assistance like ChatGPT, known for article condensation, data structuring, providing novel ideas and fresh content based on preexisting information, holds potential for enhancing TTI technology \cite{noauthor_lexica_nodate,naughton2023chatgpt,wang2023chatgpt}. Particularly in game character creation, these models can by organizing input information and interpreting generated graphic material. This capability not only aids in managing and synthesizing vast amounts of data related to character attributes and storylines but also provides valuable inspiration for AI tools designed for game character creation and interaction. 

\subsection{Human-AI Collaboration for Character Design}

Artificial Intelligence (AI) generation tools have found applications across various industries, including business, education, healthcare, and content creation \cite{fui2023generative,zhou2024eternagram}. Recent works explore the integration of AI models in creative workflows, such as artistic design, game development, and creative writing \cite{ippolito2022creative, kreminski2022unmet, yang_ai_nodate}.


In creative writing and design, tools like TaleBrush \cite{chung_talebrush_2022}, VISAR \cite{zhang2023visar}, FashionQ \cite{jeon_fashionq_2021}, and CodeToon \cite{suh2022codetoon} leverage AI for collaborative story creation and visual sketching, contributing to the creative process. Despite these advancements, there's a notable gap in using AI for character design. While tools like CharacterChat \cite{schmitt2021characterchat} support fictional character creation, they lack real user involvement in character image and profile generation \cite{schmitt2021characterchat}.


GenAI has great transformative potential in the field of human-AI co-creation processes \cite{muller_genaichi_2022,han2024teams,sun2022bringing}. The collaborative design process is categorized into stages like Q\&A, Wandering, Hands-On, and Camera-Ready \cite{hwang_too_2022}, and Spoto's framework identifies seven potential actions in mixed-initiative co-creation~\cite{spoto2017library}. These models guide the behavioral patterns of human-AI collaboration. AI collaboration tools encompass editors, transformers, blenders, and generators, all of which contribute to fostering this creative process \cite{hwang_too_2022}. Various studies, such as Scones and DuetDraw, provide insights into AI co-design tools, guiding Sketchar's interaction patterns \cite{huang2020scones,oh2018lead}.

 \added{Recent research has introduced new perspectives and solutions for content creation across various domains by combining large language models (LLMs) with generative image AI. For instance, Ray et al. utilized Stable Diffusion and ChatGPT to create speculative designs visualizing potential future scenarios related to climate change \cite{lc2023speculative}. This integration of technologies has been used to inspire creativity in the early stages of creation. DesignAID aids designers in accessing creative inspiration \cite{cai2023designaid}, CharacterMeet assists writers in developing story characters through open-ended dialogue and visual representation \cite{qin2024charactermeet}, and Text2AC helps animators bulk-generate visual appearances for 2D game characters \cite{sun2024text2ac}. However, current AI tools primarily extend functionalities from the perspective of designers, with limited exploration of how individuals without art skills, such as planners, perceive and utilize these tools. Additionally, there is insufficient research on how AI-enhanced TTI technologies facilitate expression and communication in the design process. Current virtual character creation tools often focus solely on either the appearance or the personality of characters, whereas in real character creation, these aspects are inseparable. Sketchar also aims to comprehensively consider multiple aspects of game characters to better meet the needs of game design.}

In summary, the evolving trend of applying AI in design processes aims to improve communication among participants. However, existing AI tools focus on image generation and lack the creation of text to aid collaboration, and the creative process is still artist-driven or collaborative. In addition, game character design is a relatively new field, and no tools have yet explored how to facilitate communication between game designers and game artists in a working scenario. Sketchar innovates by facilitating non-instantaneous collaboration in game character design, allowing designers to express ideas through generated characters from stories and style descriptions.


\section{Tool Design}\label{sec:ToolDesign}

\subsection{Preliminary Study}

\begin{figure}[htbp]
 \vspace{-0.3cm}
  \centering
    \includegraphics[width=.99\linewidth]{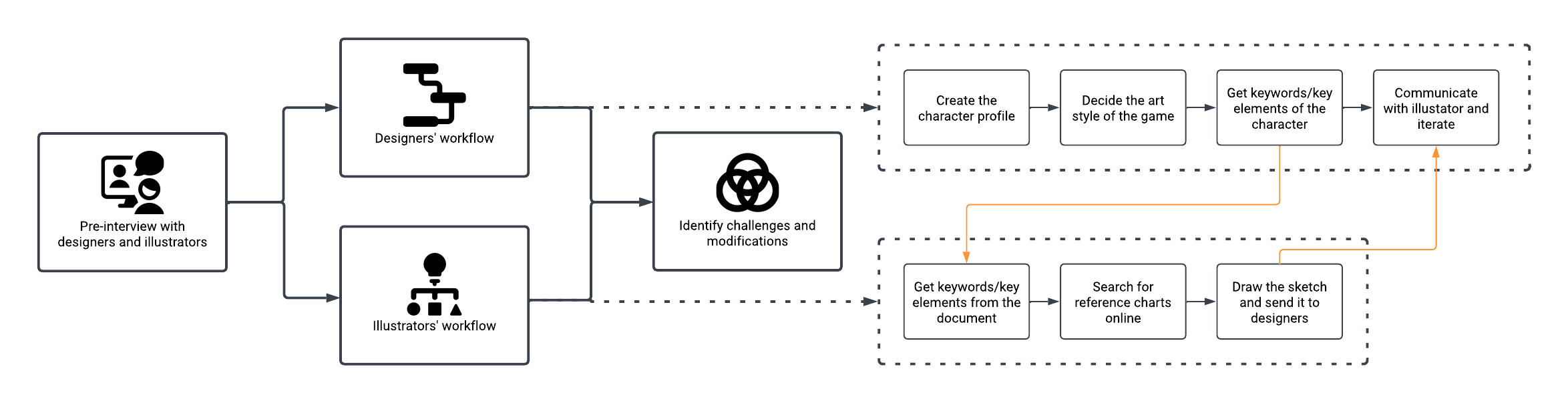}
  \vspace{-0.3cm}
  \caption{Interview Flowchart: We conducted interviews with both designers and illustrators to elucidate the workflow and collaborative pathways employed by both parties during character design.
  }
 \label{firstinterview}
  \vspace{-0.36cm}
\end{figure}

We conducted formative interviews with 10 game design professionals to understand the general workflow of character design, challenges in communication between designers and artists, and the potential support of GenAI tools for addressing these challenges. The process is shown in Figure \ref{firstinterview}.

\subsubsection{
Interviews with game designers and artists}

As shown in Table \ref{preinterviewees}, all 10 game design professionals (3 females and 7 males) studied game design, or are working in a game company. Our interviews were all conducted online. The interviews were audio-recorded and transcribed for analysis.

During the interview, we asked about (1) their general workflow when doing character design, (2) the difficulties and challenges they encounter in the design process, (3) the way designers and artists communicate and collaborate, and (4) their attitudes towards GenAI tools and the potential use scenarios.

After the interview, we applied thematic analysis and keyword coding to analyze the interview transcripts.  Following established open coding protocols, three study team members first conducted a round of independent initial coding of interview transcripts \cite{lazar2017research, braun2006using}.  Then they created a common code book through discussion. Using this code book, a thematic analysis was performed on each interview to uncover emerging topics. The research team collectively reviewed the coding outcomes to identify high-level themes.

\begin{table}[hb]
    \vspace{-0.24cm}
    \small
    \centering
    \caption{Demographic Information of Participants in the Formative Interviews.}
    \vspace{-0.2cm}
    \label{preinterviewees}
    \resizebox{\textwidth}{!}{  
        \begin{tabular}{c|c|c|c}
            \toprule 
            ID & Gender & Team Roles & Experience and Position with Character Design\\
            \midrule
            R1 & Male & Illustrator & Art Planning in a medium-scale game company.\\
            R2 & Male & Designer & Game Designer in a large game company.\\
            R3 & Male & Illustrator & Ex-Intern of prominent game companies, worked on character creation. \\
            R4 & Female & Illustrator & Master’s Student of Game Design worked on character design for an indie game. \\
            R5 & Female & Illustrator & Indie Artist, Animation student, worked on independent character creation. \\
            R6 & Male & Illustrator & Indie Artist, Animation student, and Ex-Intern in several game companies.\\
            R7 & Female & Illustrator & Lead Artist of a game project. Several experiences working as a game artist. \\
            R8 & Male & Illustrator & Professor in an art university, art consultant for indie games.\\
            R9 & Male & Designer & Game Designer in a prominent game company.\\
            R10 & Male & Illustrator & Student started doing projects for half a year, working on character design. \\
            \bottomrule
        \end{tabular}
    }
    \vspace{-0.36cm}
\end{table}

\subsubsection{Findings regarding game character design workflow}

Game character design ideation process of the designer side: In the preliminary study, our interview of game designers illustrated key phases of the character design workflow from the designer side.

\textbf{Create the character profile: }
We found that game designers would define "\textit{essential game elements, including the overarching narrative framework and core gameplay mechanics}" (R2, R9).  This initial contextual backdrop, often seen as \textit{"the precursor to detailed character development, lays the foundation for creating comprehensive character profiles"}(R2).

\textbf{Decide the art style of the game: }
In game design, "\textit{selecting the overarching artistic style}" is a pivotal creative process (R1). This decision involves negotiating various visual elements and serves as a conduit for expressing the game's theme and mechanics. Importantly, the "\textit{chosen style remains open to refinement through collaboration with illustrative team members}" (R7, R9).

\textbf{Get Keywords/key elements of the character: }
In the game's setting, "\textit{designers need to refine character keywords to align with the narrative, aiding communication with artists}" (R9).  These keywords include "\textit{core attributes and motivations, facilitating visual translation}" (R1, R6) for cohesive character integration.

Game character design ideation process of the illustrator's side: In the preliminary study, three main phases of the workflow from the illustrator's side have been identified according to the interview.

\textbf{Get keywords/key elements from the document: }
Upon receiving keywords from designers, illustrators engage in collaborative discussions with the design team.  These exchanges provide a platform for mutual exploration and alignment, fostering an understanding of the character's conceptual essence. "\textit{If both parties agree on the identified keywords, these refined descriptors solidify, serving as a definitive characterization blueprint}"(R6, R7). 

\textbf{Search for reference charts online: }
Illustrators usually begin by searching for reference images online, guided by the established character keywords.  For example, R4 noted that \textit{"I always use some websites like Pixiv, Pinterest, and Artstation to find a reference image for the specific keyword"}.   These reference images serve as visual touchstones, aiding illustrators in cultivating an aligned visual interpretation of "\textit{the character's attributes, demeanor, and thematic essence}"(R4, R6). The rationale for this approach appears to be to align the character to established standards for particular keywords.

\textbf{Draw the sketch and send it to the designers: }
Based on an initial understanding of the character, illustrators embark on the creation of preliminary sketches. These sketches are subsequently shared with designers, initiating a collaborative iteration process. Designers provide feedback and insights that reflect their vision and align with the character's established keywords. 

In the practical workflow, both the designer and illustrator undergo several rounds of discussion and iterations, with the designer ultimately confirming the character's suitability for integration into the game development process.

\subsubsection{Preliminary Findings: Challenges.}
\textbf{Communication difficulties amongst designers and artists due to divergent backgrounds and vocabularies} are reported to be their biggest challenge. In the workflow, team communication is important for “\textit{collaboration with game designers and visual artists to ensure that character design aligns with the overall vision.}”(R4)  However, “\textit{...there can be noticeable challenges in communication between the art team and the game designers. The difference in thought processes between artists and game designers can sometimes make communication difficult.}”(R6). For example, in the process of designing a non-player character (NPC), game designers might express some requirements, such as, \textit{"I envision a character with a cool demeanor, serving as a hacker, and sharing a classmate relationship with our protagonist."}(R2). The text-based communication approach here can be indirect, leading to oversight over subtle details like character looks and expressive personality that are challenging to articulate verbally. This may, in turn, result in undesirable designs and necessitate additional revision iterations. “\textit{sometimes there can be distortion or loss of information. It depends entirely on the understanding and expertise of the art team. Only if they are experienced can they quickly grasp what you couldn't convey.}”(R3).  Moreover, “\textit{when all the art assets from different artists come together to form the entire game art package, there may be slight deviations from the original elements defined by the game designer or the lead artist.}”(R7).

\textbf{Challenges in innovating and diversifying designs based on inspirational insights} were also reported. "\textit{I often fail to find the highlights and inspiration in my designs}" (R9), and "\textit{character personas are being misused; it is becoming increasingly difficult to come up with innovative and engaging character stories}" (R5). In addition, character designers described difficulties in expressing creative intentions: "\textit{I have an awesome image of a character in my head, but I don't know how to express it}" (R2), "\textit{I often have to spend so much time to find a reference image of the costume pattern I want}" (R10). Some game designers struggle with how to make the characters more vivid: \textit{When creating some RPG games, I really can't imagine how the NPC will react to interact with the main character (R9)}. 

To address the challenge, game designers reported using the following adaptations in their character design workflow.

\subsubsection{Preliminary Findings: Adaptations.}
\textbf{Keyword refinement based on character design document} was reported by respondents in the game industry. Generally, they will “\textit{extract the elements I(they) want to use, refine and integrate them, then sketch and outline the silhouette.}”(R8)  The keywords are always “\textit{Art style+characteristic.}”(R4) The characteristics will be based on “\textit{certain symbols and imagery.}”(R6) For example, when thinking about a "cool boy," they will ”\textit{think about his appearance, such as him wearing a military-style outfit.}”(R6). However, it is difficult to accurately summarize keywords. Experienced game developers tend to rely on their “\textit{inherent impressions}”(R6) and “\textit{artistic knowledge}” in this endeavor (R7).

\textbf{Using reference images enhances communication between game designers and artists}. “\textit{Sometimes, when you imagine something, it becomes difficult to describe it properly. This is especially true for original characters that have never been created before. It is challenging to describe them verbally, so finding references becomes crucial.}”(R3) Respondents reported that in their previous working experience, communicating with reference images was efficient for building a common comprehension of the characters within the team. However, the character images desired by developers may not be available as precedents online. This leads designers to “\textit{find similar designs and ask the artist to combine them or find more designs from different sources.}”(R3), which leads to more needed communication and less efficient workflows.


\subsection{Design Goals}
We endeavored to design and develop a human-AI collaborative system for facilitating character design while removing communication barriers between designers and illustrators. Drawing insights from the literature review and preliminary findings, we distilled our findings into the following design goals:

\textbf{DG1: Enhancing Keyword-Driven Character Image and Profile Generation.} To address the pain points of game designers, we aim to empower the designer's workflow, allowing for more efficient and creative character concept development. Keyword-driven generation can potentially streamline the character design process while allowing for coherent and engaging image results.

\textbf{DG2: Constructing Character Sharing Channels as Effective Communication Bridges.} Preliminary studies revealed the need for effective communication conduits between game designers and illustrators. Recognizing the critical role of cross-disciplinary communication in the collaborative process, we focused on creating structured mechanisms for sharing insights, feedback, and design requirements. The channel is intended to facilitate the exchange of ideas between stakeholders involved in character design, fostering a more cohesive creative workflow.

\textbf{DG3: Brainstorming Character Properties and Inspiring Creative Ideas by using GenAI.} We hope to apply GenAI to inspire designers to rapidly come up with ideas during character design. We hope that during this process, designers will be given impetus to imagine what characters look like, evaluate several alternatives, and be in a position to direct illustrators to emphasize key personality and relationship properties of the characters being developed.

\begin{figure}[htbp]
\centering
\includegraphics[width=1.05\linewidth]{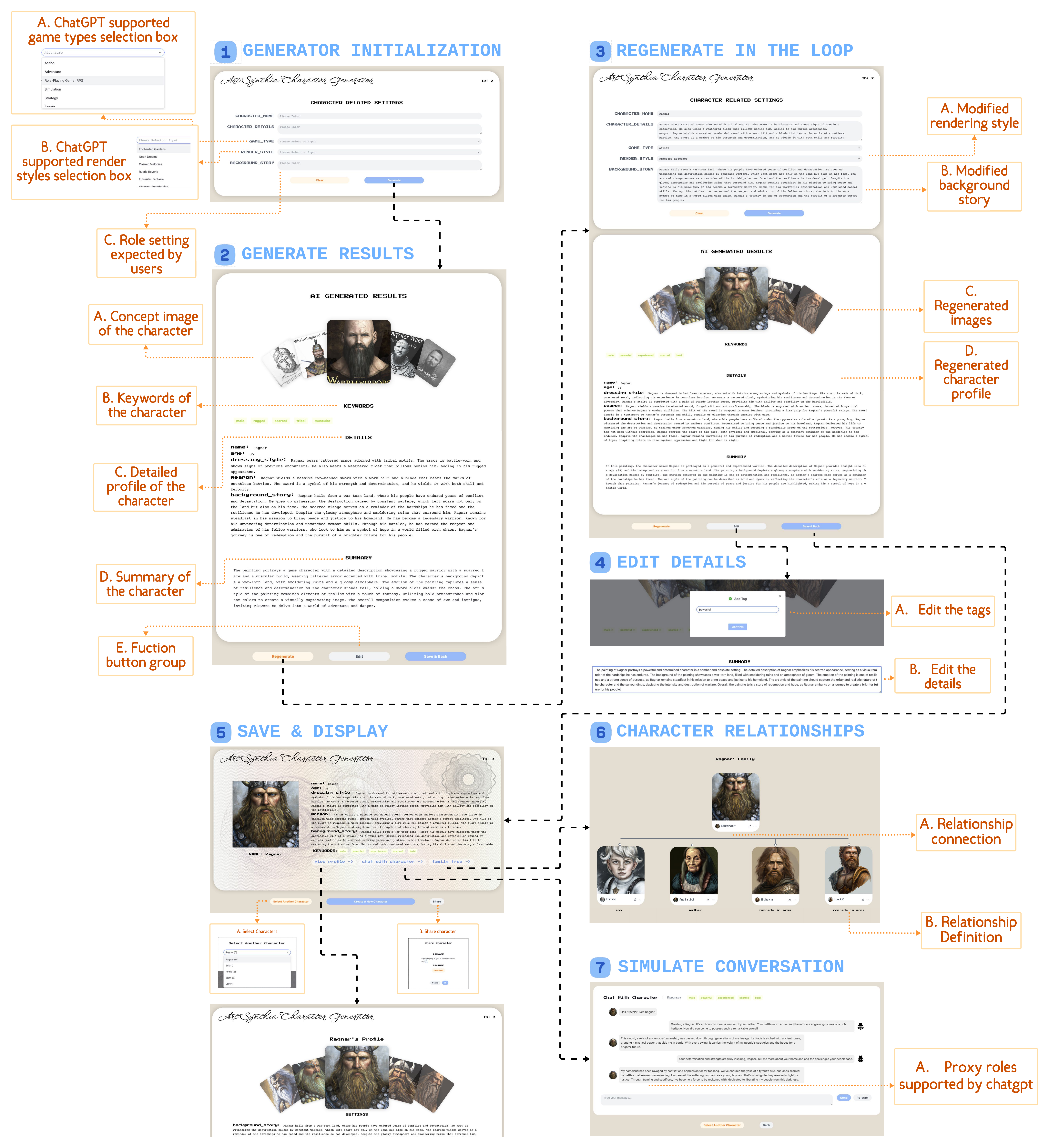}
\caption{Sketchar walkthrough. Step 1: Initialization - Users input their design intentions in the initial state. Step 2: Result Generation - ChatGPT provides users with prompt suggestions and potential images. Step 3: Modification and Regeneration - Users can modify the generator's information, and ChatGPT produces revised results. Step 4: Detailed Editing - Users have the freedom to edit character-related details. Step 5: Displaying Results - The outcomes are presented in the form of an ID card. Steps 6 and 7: Sharing with Others, Conversing with Generated Characters, and Constructing Character Lineage Trees.}
 \label{sketchar-workflow}
  \vspace{-0.36cm}
\end{figure}

\subsection{Example Scenario}

In this section, we provide an illustrative example of how Sketchar operates in a practical context. Consider a game character designer tasked with creating \textit{a brave warrior protagonist from a war-torn land}. Faced with the challenge of depicting specific attributes and appearances of characters, she opts to utilize Sketchar to organize her thoughts and gather reference images, as shown in the Sketchar walkthrough Figure \ref{sketchar-workflow}.

\removed{\textbf{Initialization.}}\added{First, }she initiates the process by entering her design intentions into the tool's initial state in Figure \ref{sketchar-workflow}.1. She can specify key traits, attributes, or contextual details related to the character's persona. \removed{\textbf{Result Generation.}}\added{Second, }upon inputting her design intentions, she interacts with the Sketchar create page, which, in turn, provides her with prompt suggestions and potential images in Figure \ref{sketchar-workflow}.2. Sketchar assists her in generating initial ideas and visual references aligned with her character concept. \removed{\textbf{Modification and Regeneration.}}\added{Third, }in Figure \ref{sketchar-workflow}.3, the designer has the flexibility to modify the information provided by the generator. In response to her adjustments, Sketchar regenerates results to fine-tune the character concept. \removed{\textbf{Detailed Editing.}}\added{Forth, }as the character concept evolves, the designer retains full control over the detailed editing process. She can further customize and refine character-related specifics such as appearance, keywords, personality traits, and background narrative in Figure \ref{sketchar-workflow}.4. \removed{\textbf{Displaying Results.}}\added{Fifth, }following the iterative design and editing phases, the tool presents the outcomes in the form of an ID card in Figure \ref{sketchar-workflow}.5. This concise representation serves as a visual reference for the designer's work.\removed{\textbf{Sharing with Others. }} \added{Sixth, }the designer can take her design process a step further by sharing the generated character concept with collaborators or peers. \removed{\textbf{Conversing with Generated Characters and Constructing Character Lineage Trees.}}\added{Seventh, } In Figure \ref{sketchar-workflow}.6 and Figure \ref{sketchar-workflow}.7, when the designer shares the generated character concepts with her collaborators, they have the option to engage in conversations with the characters generated by the tool to gain more inspiration. Moreover, they can explore the possibility of constructing a character family tree, tracing the evolution and relationships between different character iterations, and potentially gaining deeper insights into their personas.






\subsection{Technical Details}

\begin{figure}[htbp]
  \centering
    \includegraphics[width=0.9\linewidth]{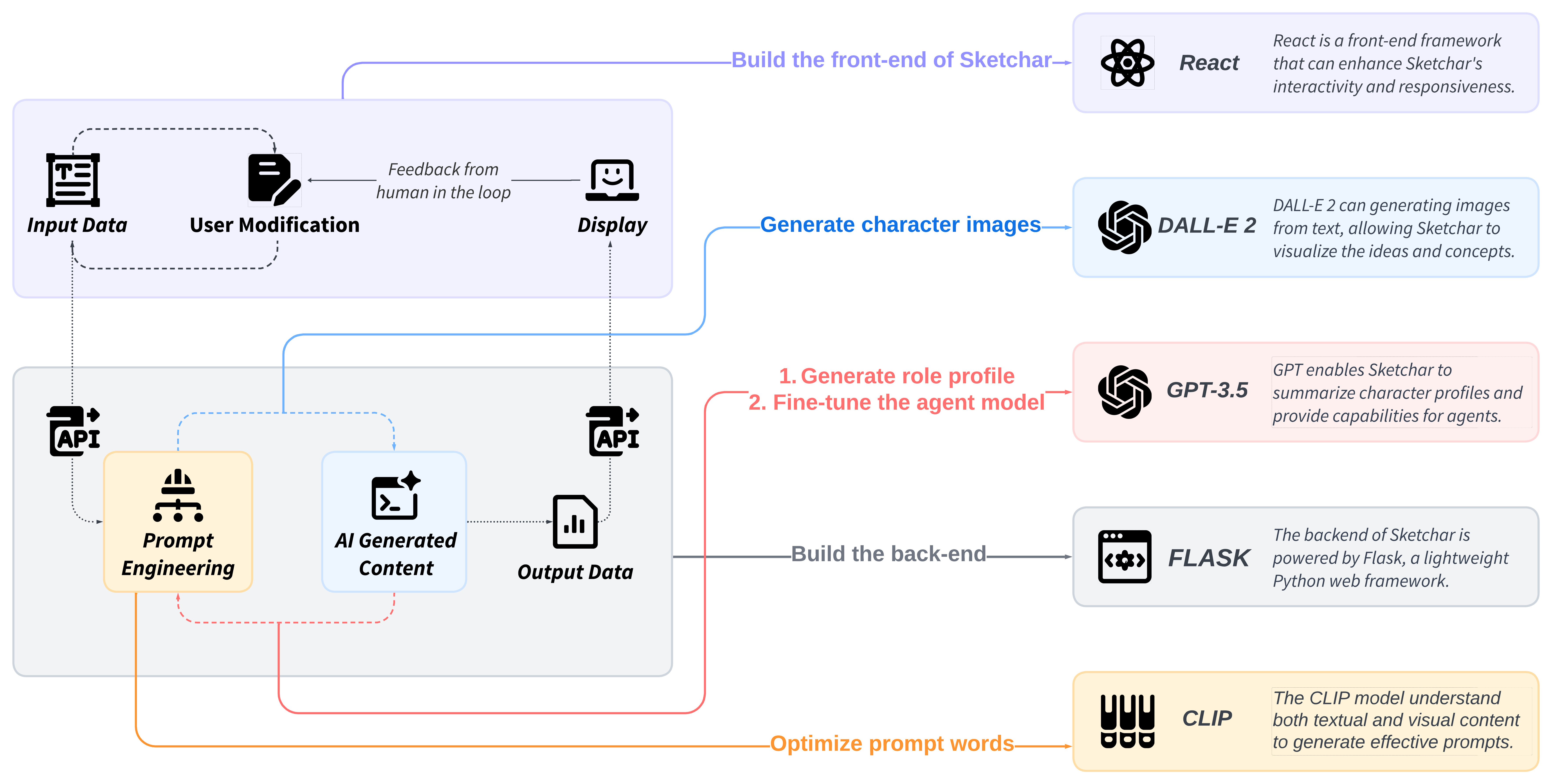}
    \vspace{-0.24cm}
  \caption{\textbf{System Architecture: } This system is devised to integrate robust artificial intelligence models(DALLE and GPT-3.5) into a web-based framework with CLIP and prompt engineering. \added{The communication parts are represented by dashed lines, while solid lines introduce the adopted technical solutions and their purposes.} }
 \label{architecture}
  \vspace{-0.5cm}
\end{figure}

As illustrated in Figure \ref{architecture}, the architecture of Sketchar \removed{encompasses an interactive web application front-end implemented in React. The back-end leverages the Flask framework to communicate with the Large Language Model (LLM) and access the MSDB database. LLM interacts with the application using the OpenAI API key, including ChatGPT and DALLE. To be specific, we have established routes and pages for various functionalities supported by Sketchar. }\added{comprises an interactive web application frontend by React and a backend utilizing the Flask framework. In the front end, users can modify the input and AI output results to achieve a human-in-the-loop co-creation effect. The backend employs Prompt Engineering to interact with ChatGPT and DALLE applications and accesses the MSDB database for data storage and management. Specifically, we have established routes and pages to support the various functionalities of Sketchar.}

\vspace{-0.36cm}


\subsubsection{Hierarchical Content Generation with Prompt Engineering}
\begin{figure}[htbp]
  \centering
    \includegraphics[width=0.99\linewidth]{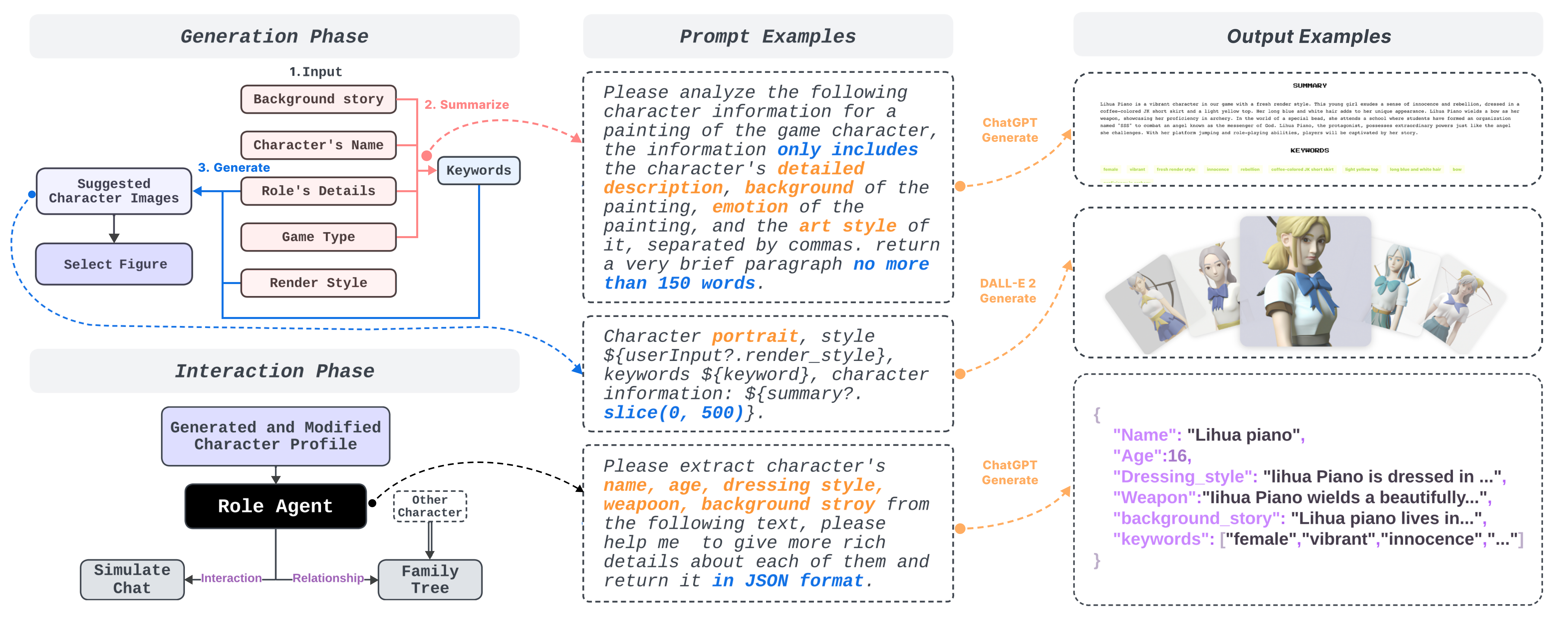}
    \vspace{-0.36cm}
  \caption{\textbf{Prompt Process Flow:} \removed{(1) Left: Demonstrating the functionalities of the Sketchar agent alongside an example of the Role Agent. (2) Middle: The generation and interaction phase encompasses a structured sequence of steps and procedures that guide the handling and interpretation of user-provided proposals. (3) Right: Presenting certain prompt words along with their corresponding steps.} \added{(1) Phases: In the generation phase, character profiles will be obtained in the order of input, summary, and generation. In the interaction phase, users can chat with the agent and build a family tree. (2) Examples: Presenting certain prompts along with their corresponding outputs. }}
 \label{prompt-engineer}
    \vspace{-0.36cm}
\end{figure}

Our tool employs a hierarchical content generation approach, supported by prompt engineering techniques, which are shown in Figure \ref{prompt-engineer}. \added{Hierarchical content generation means that the content is generated through a structured, multi-layered process, where each layer builds upon the previous one.} Offering a comprehensive overview of the pipeline, we utilize input parameters like "name," "role details," "background story," and "game type" in the first layer. In this layer, ChatGPT generates summarized integrated information using the prompts from the first one in the Prompt Examples of Figure \ref{prompt-engineer}. Progressing to the second layer, the LLM generates keywords based on the previously summarized integrated information. In the third layer, we provide both the generated keywords and "render style" and "role details" from the input to DALLE as prompts, resulting in the creation of suggested images.
Additionally, it is worth mentioning that we incorporate detailed prompts, such as "no more than 150 words" and "name, age, dressing style, weapon, background story," to shape the content and format of feedback, promoting a more structured response. In summary, this methodology empowers the tool to comprehend and address user inputs in a coherent and contextually relevant manner.

\vspace{-0.2cm}

\begin{figure}[htbp]
\vspace{-0.2cm}
  \centering
    \includegraphics[width=0.99\linewidth]{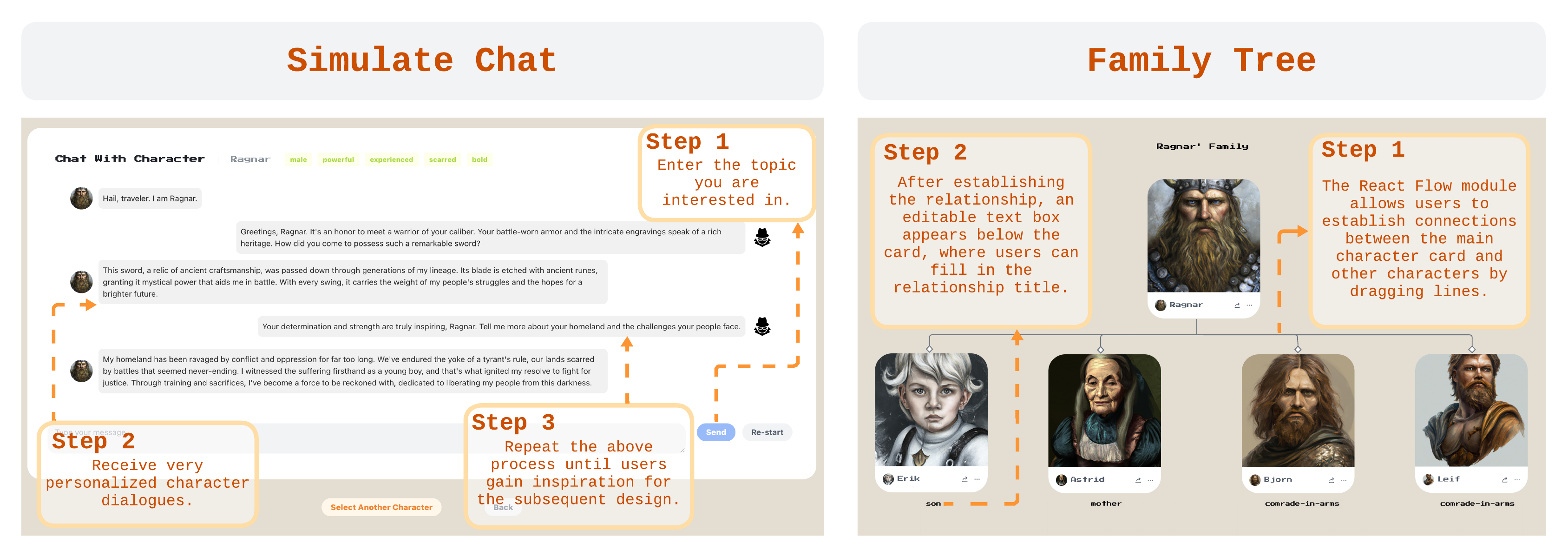}
  \vspace{-0.36cm}
  \caption{ \textbf{Simulate Chat Workflow: } Step 1, Enter the chat contents. Step 2, Receive a personalized answer. Step 3, Repeat the process. \textbf{Family Tree Workflow: } Step 1, Utilize the ability of React Flow Module to build relationships by dragging lines. Step 2, Fill in the relationship name below the card.}
 \label{agent}
  \vspace{-0.36cm}
\end{figure}

\subsubsection{Human-in-the-Loop (HITL) for Character Creation and Editing}
Our tool integrates a Human-in-the-Loop (HITL) approach within the character creation and editing process, as shown in Figures \ref{sketchar-workflow}.3 and \ref{sketchar-workflow}.4. As depicted in Figure \ref{architecture}, this approach places users at the core of character development, empowering them as active collaborators in guiding and refining AI-generated content. In terms of the technical process, by hierarchically generating content outcomes, we utilize keywords and associated content information as input for image generation, producing five reference images for character depiction. To facilitate iterative refinement based on user feedback, users can manually edit content, select the image, and repeat the aforementioned process to ensure that the resulting character profiles harmoniously resonate with their creative vision.

\subsubsection{Inspire Creativity by Customizing AI Agents and Character Family Trees}
We further implemented customization of the AI agent and a view of inter-relatedness among multiple characters through a role family tree, whose interaction process can be seen in Figure \ref{agent}. Specifically, the agent is further processed from the generated game character profile. As shown in the "Role Agent Example" in Figure \ref{prompt-engineer}, we use JSON files as fine-tuning content, allowing the role agent to utilize the language understanding, personality disclosure, and context preservation capabilities of the LLM, thereby providing users with a seamless role-playing experience characterized by participation and authenticity. Moreover, Sketchar facilitates the linkage of multiple characters' relationships. Designers can connect these relationships to construct a family tree, assisting in the organization of character relationships. While inspecting a character, illustrators can explore associated nodes, expediting access to character information.

\section{User Study}\label{sec:UserStudy}
\begin{figure}[htbp]
 \vspace{-0.5cm}
  \centering
    \includegraphics[width=.85\linewidth]{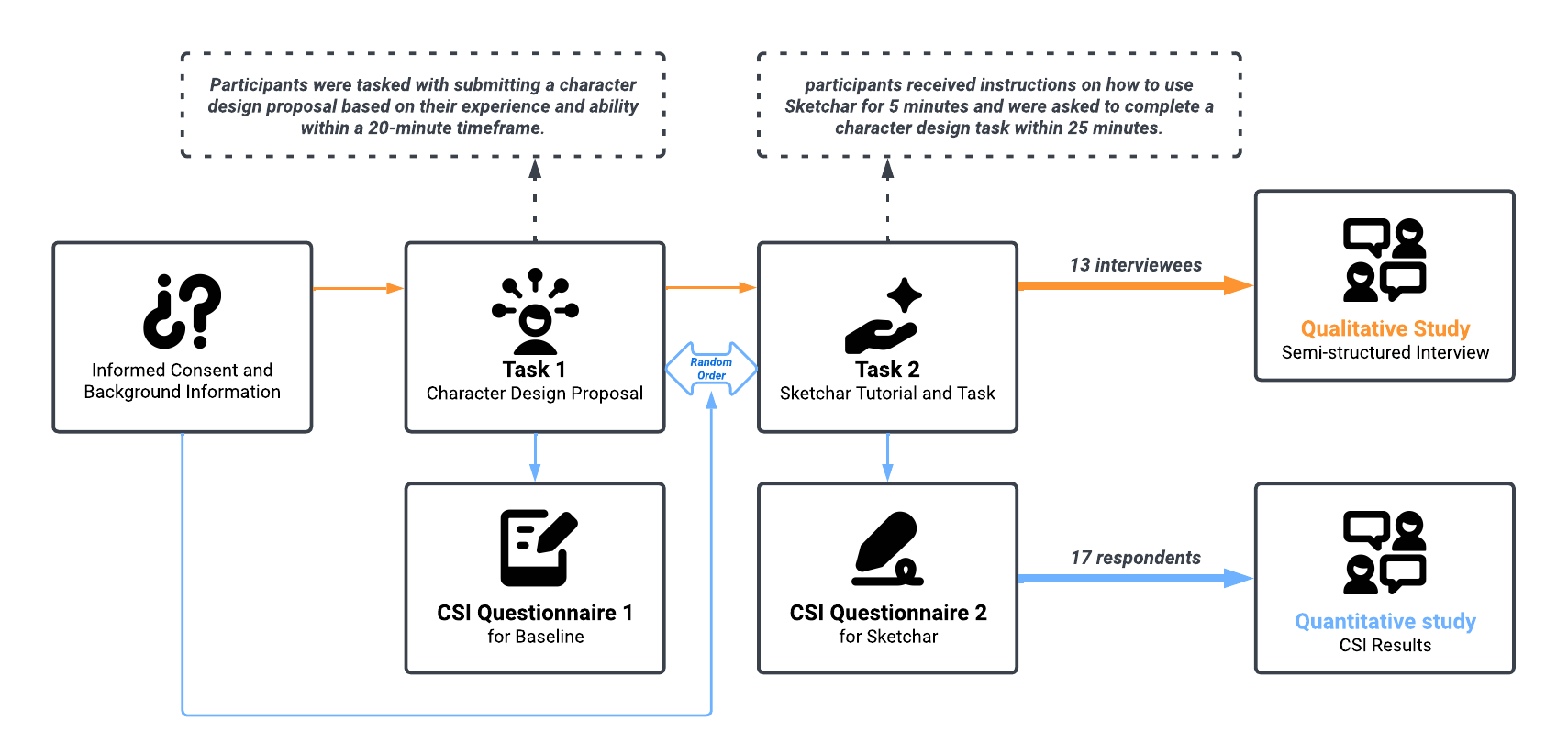}
  \vspace{-0.36cm}
  \caption{\textbf{Qualitative Study Pipeline:} (a) Commence the experiment by introducing it and obtaining informed consent from participants. (b) Allocate 20 minutes for the completion of Task 1. (c) Provide instructions for Sketchar and allow 25 minutes for the completion of Task 2. (d) Conduct a 10-minute semi-structured interview. \textbf{Quantitative Study Pipeline:} Perform the same tasks as in qualitative research, with the order of the two tasks randomized. Additionally, fill out the corresponding CSI questionnaire after each task.}
 \label{secondinterview}
  \vspace{-0.2cm}
\end{figure}

We conducted a qualitative study involving a cohort of 13 participants to examine the process by which they work with Sketchar in the character design process. We also conducted a quantitative study with 17 respondents, comparing a control (baseline) task of sketching-only workflow vs. the Sketchar rapid prototyping workflow. Details are shown in Figure \ref{secondinterview}.

\subsection{Qualitative Study on Processes of Character Design and Use of Sketchar}
\subsubsection{Participants}
Table \ref{interviewees} shows the 13 game designers from China (5 female, 8 male) we recruited for the study. The participants represented a range of professional backgrounds, including 5 participants from small-scale game development studios, 3 from prominent industry companies, 2 indie developers, and 3 students pursuing game design studies. More than 50\% participants reported having over three years of experience in game development, while 4 respondents had one to three years of experience and 2 had less than one year of experience. Our recruitment strategy targeted game developer communities on social media to ensure the inclusion of skilled game designers familiar with practical game character design. Each respondent received a compensation of 50 Chinese Renminbi (RMB) for their time and insights.

During the previous formative study, we discovered that artistic capabilities affect the process and outcomes of character design. Generally, game designers with artistic backgrounds always produce more detailed character designs and high-quality reference images. As a result, we incorporated a range of artistic prowess as one of the variables in our user testing: Six participants reported having strong artistic proficiency (having independently crafted original artwork for game characters), while the remaining seven participants lacked artistic background.

\begin{table}[htbp]
    \vspace{-0.2cm}
    \small
    \centering
  \caption{Demographic Information of Participants in the Qualitative Study.}
  \vspace{-0.2cm}
  \label{interviewees}
  \begin{tabular}{c|c|c|c|c}
    \toprule 
    ID & Gender & Art Skill & Length of Time in Game Industry & Type of Work\\

    \midrule
   P1 & Female & Yes & 1-3 years &  small-scale game development studio\\
   P2 & Male & No & more than 3 years &  small-scale game development studio\\
   P3  & Male & No & more than 3 years & prominent game company\\
   P4 & Male & No & more than 3 years & indie developer \\
   P5 & Male & No & less than 1 year & indie developer \\
   P6 & Male  & No & more than 3 years &  small-scale game development studio\\
   P7 & Female & Yes & 1-3 years &  small-scale game development studio\\
    P8 & Male &  No & 1-3 years &  small-scale game development studio\\
    P9 & Female  & Yes & more than 3 years & student\\
    P10 & Female & Yes & more than 3 years &  student\\
    P11 & Female & No & more than 3 years & prominent game company\\
    P12 & Male & Yes & less than 1 year & student \\
    P13 & Male & Yes & 1-3 years & prominent game company\\

    \bottomrule
  \end{tabular}
  \label{tab:statistics}
  \vspace{-0.36cm}
\end{table}

\subsubsection{Study Design}

We asked participants to imagine themselves as game designers for a company and assigned them the specific task of working on character design. The goal of the task was to effectively communicate the design to the game artists/illustrators so that they could accurately draw the character. To determine the task content, we conducted a demographic questionnaire survey in which we asked about the participants' familiarity with game styles and character types, as well as their previous design experiences. Two task themes were given: designing a master character for a cyberpunk-style, open-world game (task 1) and designing a master character for a \removed{secondary-style}\added{platformer anime} game (task 2). We aimed to incorporate as many common styles and genres as possible while considering the participants' design experience. The experiment lasted approximately 60 minutes and was conducted online via the Kook platform. The process is shown below:

\begin{itemize}
\item Step 1: Prior to the experiment, participants were given an introduction to the experiment process and were asked to provide informed consent and background information.

\item Step 2: Participants were instructed to submit a character design proposal to the artist based on their experience and abilities, as outlined in Task 1. They were given 25 minutes to complete the task.

\item Step 3: Following a short break, participants received instructions on how to use Sketchar for 5 minutes and were given 25 minutes to complete Task 2 with the use of Sketchar.

\item Step 4: After completing both design sessions, participants were asked questions about how they used Sketchar and what challenges and adaptations they encountered in a semi-structured interview conducted by the researchers. 
\end{itemize}

\subsection{Quantitative Study Comparing Design-only Baseline with Sketchar-Supported Design}
\subsubsection{Participants}
In this second study, we engaged 17 participants(9 males), aged from 21 to 27 years old (Mean = 23.47, SE = 1.87), recruited via snowballing method from different universities in China. The participant group was composed of two distinct profiles based on their artistic experience: 9 participants possessed prior experience in creative artistic fields (not less than one year), while the remaining 8 participants did not have any formal artistic background. We ensured that all participants had at least one game development experience as character designers in teams.

\begin{table}[htbp]
    \vspace{-0.2cm}
    \small
    \centering
  \caption{Demographic Information of Participants in the Quantitative Study.}
  \vspace{-0.2cm}
  \label{interviewees}
  \begin{tabular}{c|c|c|c|c}
    \toprule 
    Number & Gender & Age & Art Skill & Type of Work\\

    \midrule
   1 & Female & 21 & Yes & student\\
   2 & Male & 24 & Yes & small-scale game development studio\\
   3 & Male & 24 & No & prominent game company\\
   4 & Female & 24 & No & indie developer \\
   5 & Male & 22 & Yes & indie developer \\
   6 & Female & 27 & No &  prominent game company\\
   7 & Female & 25 & No & prominent game company\\
   8 & Male & 25 & Yes & small-scale game development studio\\
    9 & Male & 21 & No &  student\\
    10 & Female & 22 & Yes & student\\
    11 & Female & 25 & No & prominent game company\\
    12 & Female & 23 & Yes & student \\
    13 & Male & 27 & Yes & prominent game company\\
14 & Male & 22 & No & small-scale game development studio\\
15 & Female & 23 & No & prominent game company\\
16 & Male & 22 & Yes & small-scale game development studio\\
17 & Male & 22 & Yes &   small-scale game development studio\\
    \bottomrule
  \end{tabular}
  \label{tab:statistics}
  \vspace{-0.3cm}
\end{table}


\subsubsection{Study Design}
 
In the second user study, we replicated the steps employed in the first study, introducing a modification: the order of task 1 (Baseline Control Condition) and task 2 (With Sketchar) was counterbalanced by randomization. Unlike in the first study, we asked participants to complete a Creativity Support Index (CSI) survey \cite{cherry2014quantifying} to collect user feedback after each of the trials, both after task 1 and after task 2. The survey attempts to measure how well the tool supports creative activity on a scale of 0 to 10 in six dimensions. \removed{Figure \ref{csicontent} showed the content of the scale.}\added{The content of the scale can be found in the appendix.} To ensure semantic coherence, within the baseline condition, we replaced the description "the system or tool" with "the process."

Furthermore, the collaboration subscale of the CSI generally refers to synchronous collaboration with another person. However, the asynchronous collaboration that Sketchar supports between designers and artists. Considering Sketchar's design goals, we still included this subscale in our study. During the experimental process, we informed participants that the term 'collaboration' referred to cooperation with a potential artist, so they had to imagine working with an artist/illustrator on the project.

 

 
 

\begin{figure}[ht]
\vspace{0.0cm}
\centering
  \includegraphics[width=1\linewidth]{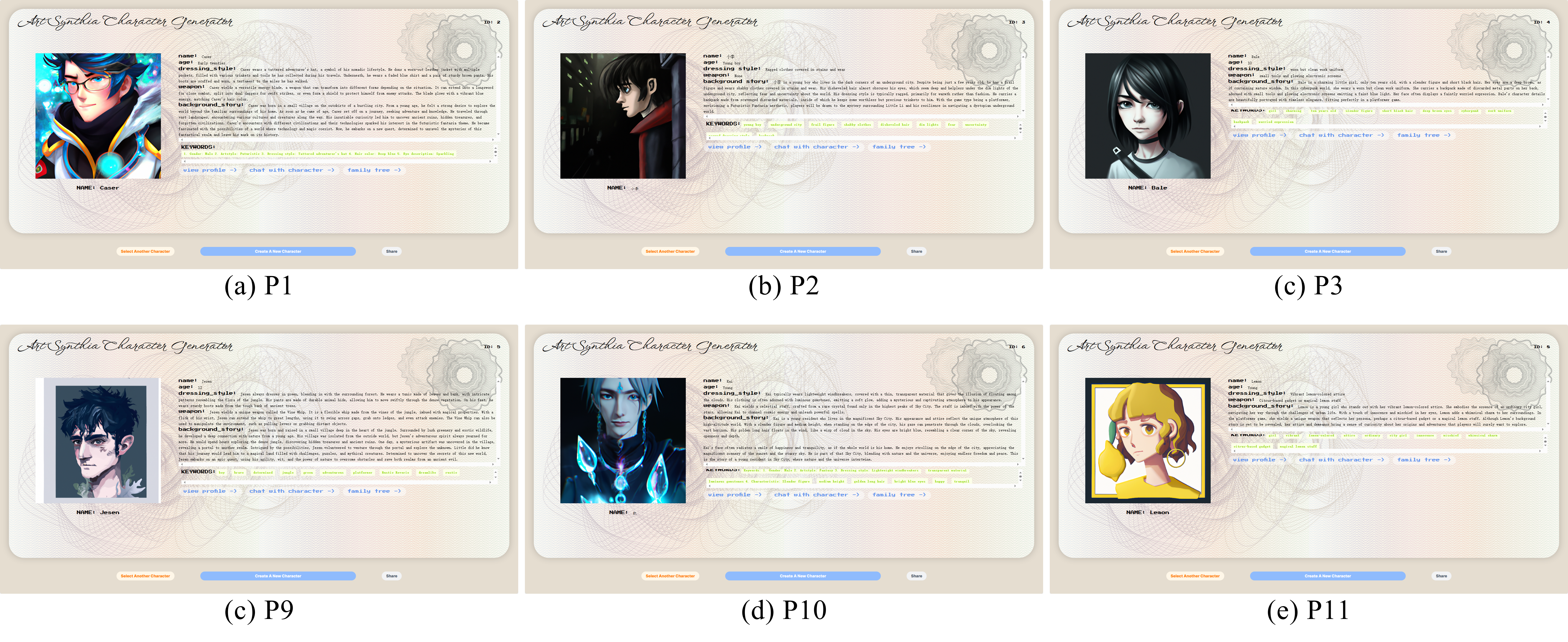}
  \vspace{-0.8cm}
  \caption{In the interviews, the interviewees received Task 2: Generating the protagonist for a fresh platform jumping adventure game. The images displayed include partial results generated using Sketchar, involving six experts (P1, P2, P3, P9, P10, and P11).}
	\label{results}
  \vspace{-0.5cm}
\end{figure}

\subsection{Qualitative Results from Study on Character Design and Use of Sketchar}

\begin{figure}[ht]
\vspace{-0.12cm}
\centering
  \includegraphics[width=1\linewidth]{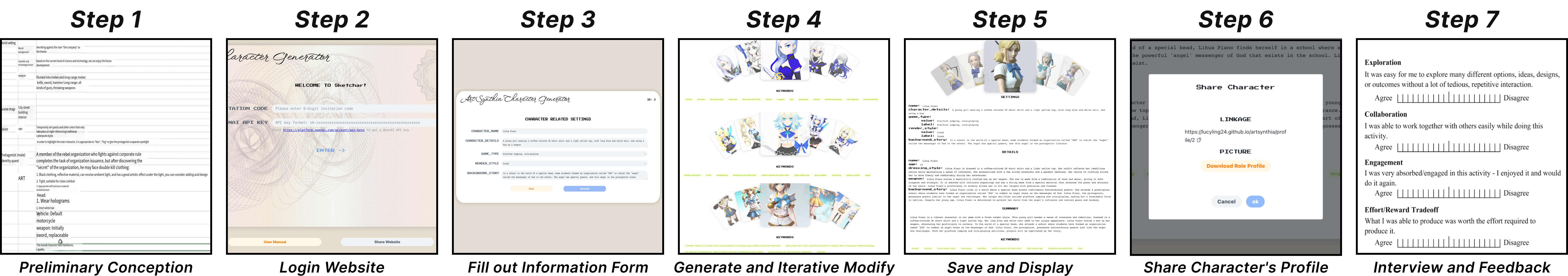}
  \vspace{-0.5cm}
  \caption{\removed{The following steps are part of the testing process, as interviewee P6 exemplifies} \added{A study process example of interviewee P6}: Step 1 involves preliminary planning and brainstorming. Step 2, the participant logs into the website. Step 3 entails completing a character form. Step 4 involves generating a character using tools and iteratively refining it. Step 5 encompasses saving and reviewing the presentation results. Step 6 downloads the character file for sharing. Step 7 entails undergoing a semi-structured interview and questionnaire survey.}
  \label{usertestworkflow}
  \vspace{-0.12cm}
\end{figure}

Drawing upon observation records and semi-structured interview data, we uncovered the workflow of how users use Sketchar (Shown in Figure \ref{usertestworkflow}). We performed thematic analysis on the initial coding from the interview data, and through collaborative discussion by all three researchers, we reached \removed{consensus} \added{a consensus} on key findings from the study.

\textbf{KF1: Artistic participant backgrounds lead to a higher demand for details of game character design.} The user testing phase elucidated a notable trend: participants with artistic skills and backgrounds tend to craft comprehensive design documents during Task 1. They also exhibited a propensity towards discontentment with the detailed output generated by Sketchar, especially in the generated reference diagram section. For example, P12 said: \textit{“The generated image is a little bit easy and lacks many details compared with the standard of my workflow”}. Without details, the image now generated \textit{“is hard to utilize in pragmatic workflows”}(P13). However, the study results from participants without artistic backgrounds reveal a different insight: \textit{“Something is better than nothing”}(P2);  \textit{“Since I don't know how to represent my design in the form of art, I thought generating some simple reference images would help me communicate my design”}(P3). 

\begin{figure}[htbp]
\centering
    \begin{minipage}[t]{0.48\textwidth}
        \centering
        \includegraphics[width=1\linewidth]{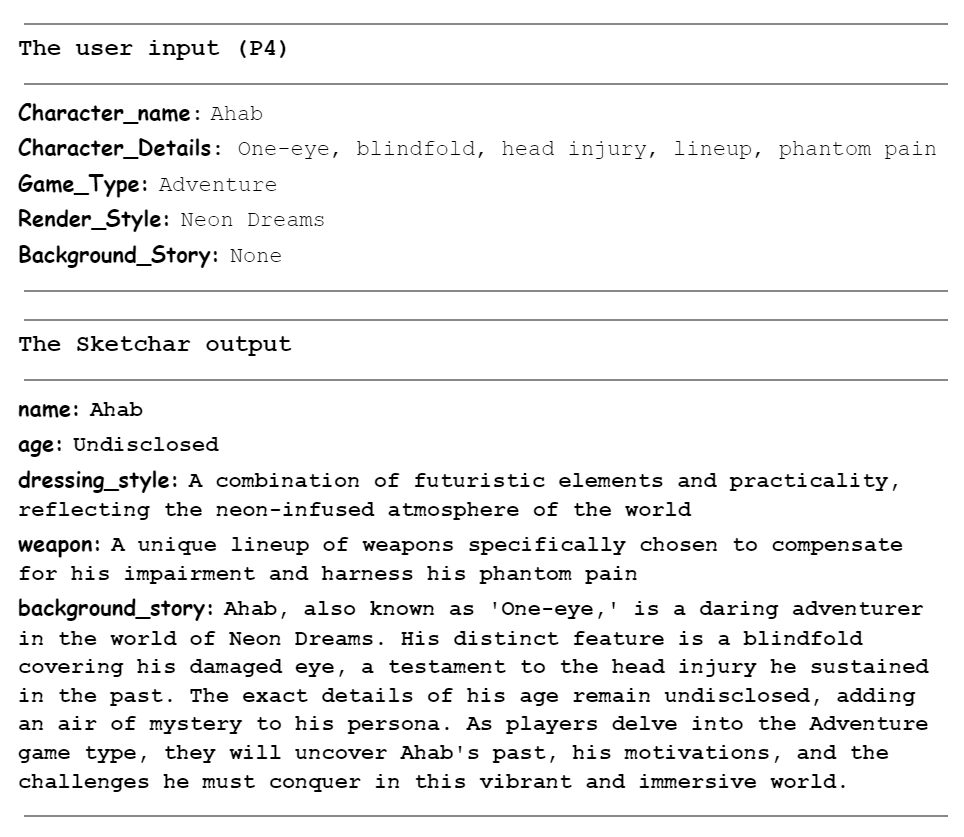}
        \label{inout}
        \vspace{-0.4cm}
        \caption{An example of import and output content from P4. He designed a master character named Ahab for a platformer anime game.}
        \vspace{-0.36cm}
    \end{minipage}
    \hfill 
    \begin{minipage}[t]{0.48\textwidth}
        \centering
        \includegraphics[width=.65\linewidth]{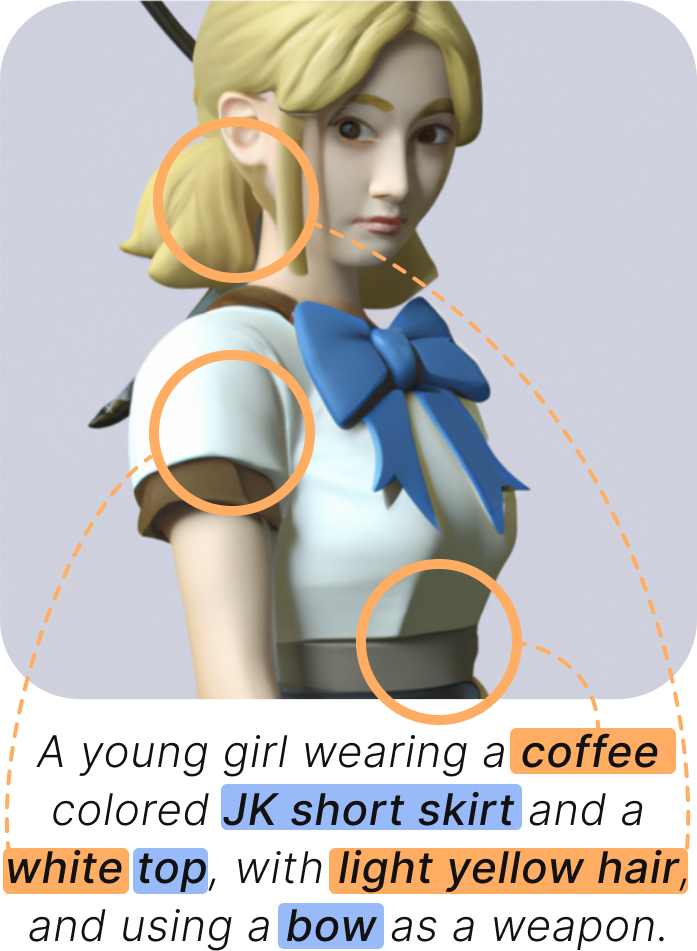}
        \caption{Correspondence \removed{graph} \added{of the image and prompt words}: Taking participant P5's results as an example, \removed{the figure maps the generated outcomes to the produced results.} \added{and mapping features of the outcome image to the original inputs.}}
        \label{fk3}
        \vspace{-0.36cm}
    \end{minipage}
\end{figure}

\textbf{KF2: Classifying and expanding character design concepts through GenAI assists game designers in expressing creative ideas, fostering refinement of design details.} Many participants noted that Sketchar allowed them to elaborate on simple and early-stage character design concepts and generate a more coherent design document.  Sketchar appears to aid them in structuring their design concepts quickly, enabling them to progress to more detailed stages. According to one participant, it \textit{"can help with basic character analysis and style generation. The text sorting and classification aspect is commendable and highly effective, particularly in swiftly organizing appearance, weaponry, and personality details"} (P11). It’s also beneficial for game designers to conduct further design work. \textit{“The generated documents can serve as the basis for my further detailed design. Meanwhile, the expanded generation of the character's worldview by this tool can also inspire me to think about the whole character's setting”}(P2).

\textbf{KF3: The generated reference images provided assistance in visually expressing design requirements.} According to the participants, they can usually quickly generate an abstract conceptualization of a character given inspiration. However, fully expressing their ideas to illustrators for design realization proves more challenging. \textit{"I understand the essence and feeling of the character, but articulating it becomes difficult, resulting in communication and dialogue misunderstandings with illustrators in the team"}(P2). While reference images can be helpful in such situations, certain issues may arise. \textit{"Merely searching for reference images cannot resolve these matters as many desired characters may not currently exist."} (P5) Participants utilized Sketchar's capacity to produce reference images based on design concepts. Although the quality of the generated images is not flawless, they reported that they can serve as a foundation for communication. \textit{ "I can inform the illustrator that I desire the face and hairstyle depicted in this picture while requesting adjustments regarding hair and clothing coloration. In this manner, the communication document commences with a visual representation to convey my preferences and concerns."}(P5) P5's results are shown in Figure \ref{fk3}.

\subsection{Quantitative Results from Study Comparing Baseline with Sketchar-Supported Design}



\begin{figure}[ht]
\vspace{-0.6cm}
\centering
  \includegraphics[width=0.8\linewidth]{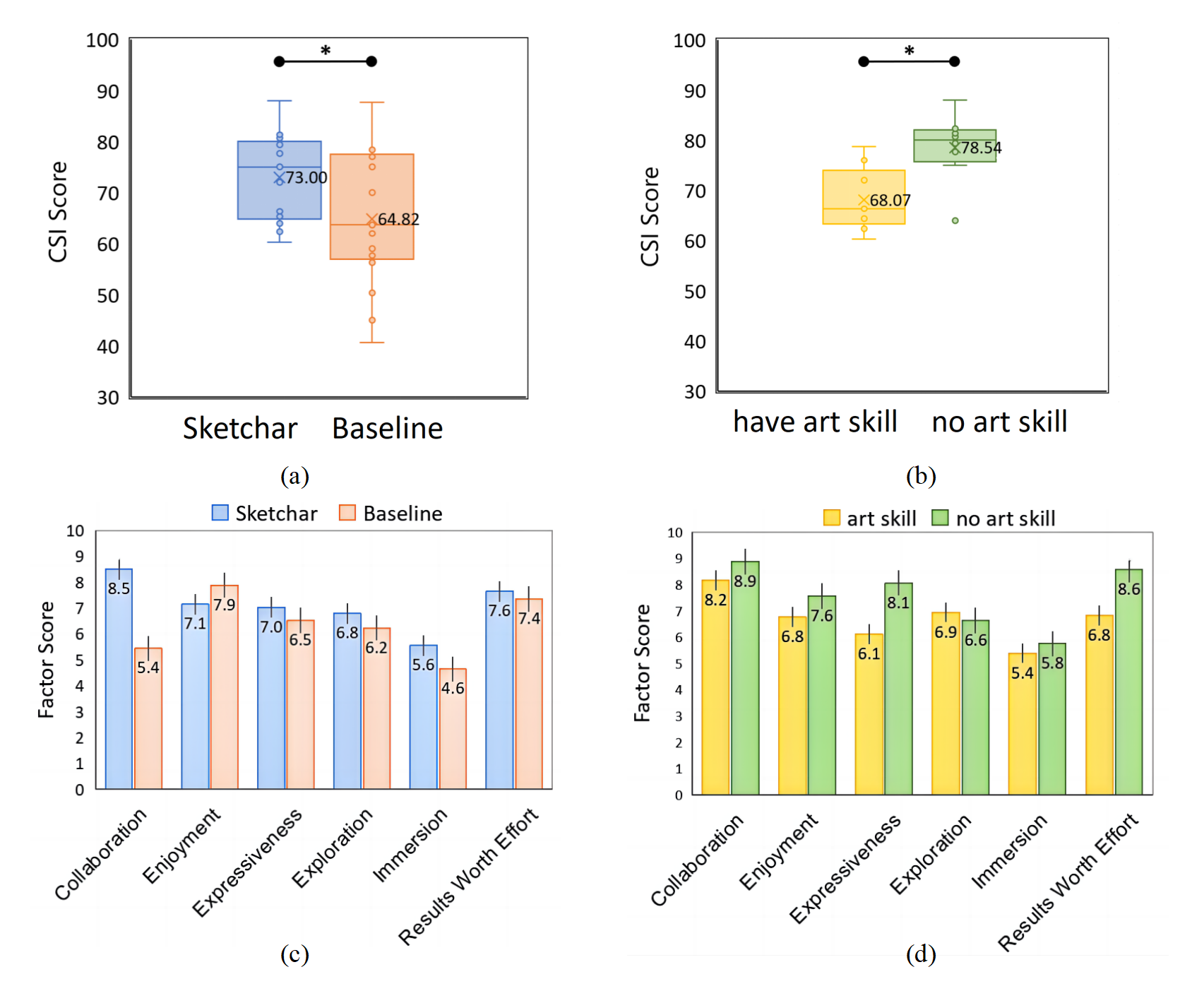}
 \vspace{-0.36cm}
   \caption{The comparative analysis of the total CSI scores and factor scores. (a) The average CSI score for Sketchar is significantly higher than the CSI score for the baseline condition (p= .028 < .05*). (b) In the group without artistic skills, Sketchar's CSI score is significantly higher than in the group with artistic skills (p= .011 < .05*). (c) The average scores in six dimensions under Sketchar and baseline conditions. Sketchar scores are significantly higher than the baseline scores in the Collaboration dimension. (d) The average scores in the six dimensions of Sketchar for participants with and without art skills. In the expressiveness and Results Worth Effort dimensions, the group without art skills scored significantly higher than the group with artistic skills.}
   \label{csifig}
   \vspace{-0.36cm}
\end{figure}

\removed{Table \ref{tab:csitotal} shows the CSI factor scores and total scores for the Sketchar and baseline conditions.}\added{Figure \ref{csifig} shows the creativity support index for different workflows and different art skill levels. First, the data did not follow a normal distribution, we performed a two-way nonparametric ANOVA, and no significant interaction was found between the variables (\textit{p} = .146 > .05, Scheirer-Ray-Hare test).} \removed{Figure \ref{csifig} shows that the CSI score for the Sketchar condition is significantly higher than that under the control baseline condition (Wilcoxon ranked sums, p =.029). }\added{Due to the non-normal distribution of differences between data pairs, we used the Wilcoxon signed-rank test to analyze the CSI score. Figure \ref{csifig} shows that the CSI score for the Sketchar condition is significantly higher than that under the control baseline condition (\textit{p} = .028).} \removed{Each CSI factor appeared to be internally consistent (Table 4, Cronbach's Alpha).}\added{The Cronbach's alpha of each CSI factor was greater than 0.7, indicating that all factors possess internal consistency. Multiple comparisons and corrections were performed on the factor scores of the different dimensions to further explore the differences between the CSI data. Tukey test showed that Immersion scores are significantly lower than scores on the other dimensions, while Collaboration scores are higher than any other dimensions.} \added{We did not statistically test each factor score. Instead, we compare them directly, which is a suggested way to compare two different creativity support tools \cite{qin2024charactermeet, cherry2014quantifying}.} \removed{Sketchar scored higher in the Collaboration dimension than the baseline condition, indicating that participants perceive the tool as facilitating cooperation between character designers and artists. Although the differences were not significant on all dimensions except enjoyment, Sketchar scored higher than Baseline control. We speculate that the higher scores in enjoyment for the baseline could be attributed to participants being more accustomed to existing tools and general workflows for task completion.}
\added{On all dimensions except enjoyment, Sketchar scored higher than baseline control, especially in the Collaboration dimension. The factor score suggested that participants perceive the tool as facilitating cooperation between character designers and artists.  For the differences in the Enjoyment dimension, it suggested that the higher scores in enjoyment for the baseline could be attributed to participants being more accustomed to existing tools and general workflows for task completion.}

We compared the CSI scores for Sketchar between participants with and without art skills \removed{using the Wilcoxon test}\added{using the Wilcoxon rank-sum test}. Figure \ref{csifig} (b) shows that participants reporting lacking professional artistic skills scored significantly higher than those reporting possessing professional artistic skills (\textit{p} = .011). Character designers without great artistic skills showed greater interest in our tool and provided more positive evaluations. \removed{Significant differences in factor scores were observed, particularly in the dimensions of Expressiveness (\textit{p} = .002) and Results Worth Effort (\textit{p} = .015).} \added{Large differences in factor scores were observed, particularly in the dimensions of Expressiveness and Results Worth Effort. The difference on these two dimensions reaches more than 2 points.}  We infer that participants with art skills, capable of creating their own characters, likely compared the outcomes generated by the tool with their own drawing results, resulting in a more critical assessment of the tool's expressiveness and cost-effectiveness. Interestingly, the scores in the exploration dimension were slightly higher for participants with artistic skills. The difference suggested that as artists, they may value the exploratory inspiration offered by the Sketcher to a greater extent.

\subsection{Expert Evaluation of the Quality of Images Co-created with Sketchar}

To evaluate the qualities of images generated from co-creation with Sketchar's, and how well they can potentially facilitate real-world communication between designers and illustrators, we engaged five illustrators from the preliminary study as expert evaluators. These experts were tasked with assessing the applicability of Sketchar's outputs as conceptual references for professional game development in comparison with their own workflows. 
We selected five distinct prompts from the former study used by designers, for which Sketchar generated ten reference images each. The evaluators then critically reviewed the quality of each image to determine its suitability as a reference in actual design workflows, much as their own created images.

\begin{table}[htbp]
    \vspace{-0.2cm}
    \small
    \centering
  \caption{Demographic Information of Experts.}
  \vspace{-0.2cm}
  \begin{tabular}{c|c|c|c}
    \toprule 
    ID & Gender & Time in Game Industry & Experience and Position\\

    \midrule
   E1 & Male & 3-5 years & Art Planning in a medium-scale game company\\
   E2 & Female & more than 5 years & Indie Artist, Animation student, worked on character creation.\\
   E3 & Female & more than 5 years & Lead Artist of a game project. Several experiences.\\
   E4 & Male & more than 5 years &  Professor in an art university, art consultant for indie games.\\
   E5 & Male & 3-5 years & Indie Artist, Ex-Intern in several game companies. \\

    \bottomrule
  \end{tabular}
  \label{tab:experts}
  \vspace{-0.12cm}
\end{table}

\begin{figure}[ht]
\centering
  \includegraphics[width=1\linewidth]{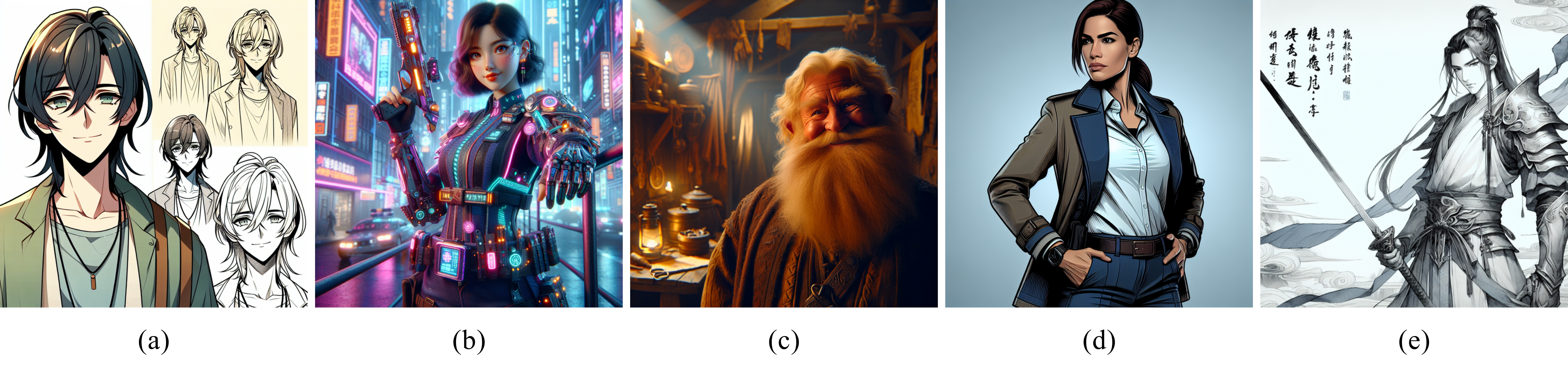}
  \vspace{-0.8cm}
\caption{Example reference images generated by participants in the study. (a) Prompt 1: A very cool boy with long hair for a 2D anime AVG. (b) Prompt 2: In an RPG game, a very cute girl in the Cyberpunk world, is a policewoman using a gun to punish crimes. (c) Prompt 3: An NPC for a DND-like game, a wise old dwarf, very warm-hearted, always smiling. (d) Prompt 4: A main character for a reasoning game, a very cool and serious girl, who is one of the greatest detectors in the world. (e) Prompt 5: In a Chinese-style RPG game, a Chinese-ink style knight, holds a sword, he looks very cool and always calm.}

	\label{experts}
  \vspace{-0.12cm}
\end{figure}

The quality of 10 images in each generated set is assessed to ensure relevance and creativity in aiding game design processes.   To achieve this, we employed a workflow where experts, familiar with the process of game character design, ranked the quality of images based on their practical utility in actual design workflows. Then they would decide how many generated images in each set are qualified as reference images in real-game character development. This process is consistent with their real workflow: illustrators would evaluate the quality of the reference images provided by the game designers to decide if they are "useful for further character development in your workflow." The experts identified an average of 6.84 images per set(ten) as relevant valuable references for game character artists. This suggests that Sketchar is capable of producing character images from keywords that would contribute to real-world workflow (Figure \ref{experts}).


 

 



\section{Discussion}\label{sec:Discussion}

\subsection{GenAI as a Support Tool for Expression and Inspiration}

It is difficult to articulate thoughts in a particular medium without professional skills. Recent research has employed GenAI to aid users in expressing their ideas \cite{chung_talebrush_2022, mirowski2023co, han2023design}. However, these studies have focused on storytelling and expression, with limited research dedicated to the type of interdisciplinary design expertise necessary in, for example, the crafting of characters for games. This domain suffers from the issue of created characters becoming stereotypes when lacking originality from their creators \cite{voimala2023creating}. This concern was echoed in our preliminary study, where participants reported difficulties in innovating and diversifying their designs. GenAI leverages large data to provide a natural language interface for idea generation, promising to be a solution for designers to overcome issues of lack of originality \cite{kwon2023understanding}, applied in domains like fashion \cite{jeon_fashionq_2021} and design sketching \cite{arora2017sketchsoup, huang2019sketchforme}.

Based on preliminary expert interviews, we designed Sketchar specifically for the creation of characters in games while addressing the most serious pain point of communication between designers and illustrators. Sketchar utilizes a keyword-driven approach to character design, which was found to be a standard that is efficient for professional game designers in our formative interviews. Previous work proposed three requirements for the Creativity Support Tool (CST) \cite{chung2022artistic}: 1) allow users to express under-constrained intentions; 2) enable the tool and the user to co-learn user expressions; 3) allow easy and expressive iteration. Firstly, Sketchar enables designers to input content freely, including common game genres and styles. This input is then mapped to associated images, assisting designers in expressing their creative intentions. Secondly, we employ ChatGPT to structure user inputs and generate coherent character profiles, aiding in the organization of designers' thoughts. Finally, Sketchar allows users to edit and regenerate character profiles, with the option to save preferred outcomes at any stage.

The CSI score findings demonstrate the efficacy of Sketchar in the character design process. Sketchar's expressiveness results are notably superior to baseline levels for persons without artistic backgrounds. The disparity may arise due to those character designers who lack artistic abilities resorting to non-customized approaches like online image searches, which would not be able to express their intentions well. Thus, those with limited artistic ability would be more likely to find Sketchar's capabilities in generating character images liberating. Conversely, character designers who possess artistic abilities may uphold more stringent criteria for the excellence of character illustrations. Qualitative insights from interviews further supported this trend. Participants with art backgrounds frequently commented on the quality of the generated images, exemplified by P13's observation: \textit{"The system is intriguing; however, I am not very satisfied with the generated images."} People with art skills seem to have more concerns about AI-generated images \cite{chung2023artinter, ko2023large}. For example, in Ko's study \cite{ko2023large}, they interviewed 28 artists to evaluate images generated by Large-scale Text-to-image Generation Models, and the results showed four key limitations of LTGMs from artists' perspective: First, their predictability limits unexpected, creative inspirations. Second, LTGMs lack the ability for deep personalization and fail to address practical design specifics, such as architectural feasibility or detailed fashion elements. Third, the reliance on text prompts restricts the expression of novel or abstract concepts, hindering the artistic process. Finally, the inefficiency and cumbersome nature of LTGMs make them less available for artists with specific visions, highlighting a gap between LTGMs capabilities and the practical needs of artistic creation. 

During the qualitative user testing interviews, designers evaluated Sketchar's character profile organization feature. During the first phases of character design, designers often integrate a wide range of various aspects and ideas (P6) and develop profiles to elucidate their design concepts. In addition, several designers observed that when they experience a lack of ideas or originality, the character profiles produced by Sketchar provided a source of inspiration. While these profiles may not serve as the final version of the character design, they appeared to provide a diversity of materials for the designers to edit. The generated images are perceived as potential concepts and style guides for characters, though the expressiveness in fine details like accessories and clothing patterns remains somewhat limited. \removed{Existing studies such as SketchSoup \cite{arora2017sketchsoup} inspire creators by inputting rough sketches and generating new sketch sets. Sketchar inspires designers by inputting textual descriptions to generate character stories and images, allowing designers to be influenced by both textual and image information. Regarding the lack of finer details, Sketchforme \cite{huang2019sketchforme} bridges the gap by generating batch images for users to choose from; SketchSoup \cite{arora2017sketchsoup} generates high-quality images through context matches for sketch alignment.}

Game designers frequently need to produce a template for a non-player character (NPC) that is then given different variations when instantiated. Sketchar has the potential to produce NPCs at scale for designers. In each generation, the tool can produce five similar character images that may serve as different variations for one particular design. This may be more efficient than manually creating each NPC since NPCs often do not require extensive diversity but rather need to align with the game’s setting. \removed{Existing studies have mainly batch-generated NPCs based on image input. One study created NPC images based on artists' blurred input images \cite{guo2022converting}, and another tool called ArtBreeder allows users to upload images of various types of characters and modify the character design images by adjusting the basic attributes.}Sketchar batch-generates characters based on textual inputs instead of image inputs to keep the style consistent and focus more on the character's personality and story. Future work is needed to extend this capability to customize the way these NPC variants can be adapted to different contexts of the level design in the game. In the future, Sketchar has the potential to alleviate the repetitive workload of character designers in the unique situation of large numbers of NPCs.

\added{In comparison to other studies on AI-assisted character design tools\cite{guo2022converting, huang2019sketchforme, arora2017sketchsoup, cai2023designaid,qin2024charactermeet, sun2024text2ac}, Sketchar can be personalized to meet the expressive needs of game character designers and facilitate collaboration and communication during the design process. For instance, we compared our tool to CharacterMeet, an AI tool that supports creative writing by building characters through dialog. Both of them utilize ChatGPT and visual AI tools to assist designers in creating characters, allowing designers to be inspired by both textual and image information. However, there are differences between them. CharacterMeet collects characters’ information through conversations and guides character development through conservation. while Sketchar organizes and iterates characters by profiles. Additionally, CharacterMeet produces 3D characters with audio and visuals for enhanced immersion, whereas Sketchar predominantly generates 2D characters with diverse art styles tailored to different game genres. Moreover, Sketchar’s versatility allows its application across different game styles, expanding the scenarios for using character design tools. Unlike CharacterMeet, which excels in detailed character development suitable for linear creative writing, Sketchar supports batch generation of game characters, meeting the needs of early-stage design and idea expansion. It is also thought to be useful for generating large quantities of characters with high similarity and simple design, such as NPCs, quickly. In our evaluation methods, CharacterMeet's participants are all experienced creative writers, while Sketchar's participants include designers both with and without artistic skills, targeting a broader group. We both collected qualitative and quantitative data through interviews and CSI scales, while CharacterMeet further encodes users' think-aloud content, providing more direct insight into their thought processes. This aspect can be incorporated into Sketchar in the future. In summary, our tool is mainly used in the game design field, focusing on facilitating the expression of ideas by designers who do not have a high level of artistic ability and generating game characters that meet the requirements of the game in a batch.}

\subsection{GenAI as a Facilitator for Iterative Design and Illustration.}

In formative interviews, we have observed that character design is often not a singular process, as illustrated in Figure \ref{firstinterview}. It typically entails iterative design of characters through character briefs, game plots, and dialogues by game designers and illustrators. While most character design tools focus on achieving visual representation of characters, such as the PokerFace-GAN \cite{shi2020neutral}, which automates the creation of neutral-faced game characters, they address the customization challenge in role-playing games (RPGs), particularly the task of generating game characters from a single photo. In our research, as we immerse ourselves in the gaming context, we also consider contextual factors for image generation, such as the game background. This stands as the primary distinction between our work and theirs.

Moreover, in the iterative process of character development, communicative virtual character agents may facilitate this process. Prior works support such claims, like CharacterChat \cite{schmitt2021characterchat}, which revolves around character agents specialized in personality classification. They introduce the Social Support Conversation (S2Conv) framework as a novel solution to address challenges faced by traditional emotion support methods. The core contribution of this research lies in establishing the MBTI1024 bank, underscoring CharacterChat's capability in personalized social support and highlighting the crucial role of interpersonal matching in enhancing support effectiveness. This is inspiring for our character design system. We believe that providing such interactive experiences through character agents can serve as a design \removed{paradigm}\added{method}. When creating new characters, enhancing interactive engagement and perceptual abilities within existing character generation systems can be achieved by assisting users in iterating content and providing additional references.

Furthermore, the character "Family Tree" feature allows designers to efficiently organize relationships between multiple characters generated by GenAI using templates. This facilitates designers and illustrators to quickly understand the relationships between various characters, which we believe also contributes to the iterative design of characters.

The study findings could serve as evidence of effectiveness. The Exploration and Result-Worth-Effort subscores in the CSI results exceeded scores in the baseline control. One explanation is that these features inspire iteration and enrichment in their character design,  providing users with exploratory ideas. Particularly, as emphasized by one participant (P5) who engaged in feedback-based conversation during trials,\textit{"That's awesome, my character is talking to me! Such a surreal feeling!"}. Participants found it exciting and somewhat surreal that their characters were conversing with them. Thus, we can infer that providing these features in Sketchar is an effective means during assisted iterative character design.

\subsection{GenAI as a Mediator for Effective Team Communication}
Several previous studies have explored the role of GenAI in collaborative creativity \cite{chung2022artist,chung2023artinter}. These studies revealed AI's potential to foster collaboration among users with different roles. \removed{For example, We-toon is an innovative communication support system that bridges the contextual gap between webtoon writers and artists during sketch revisions. It enables writers to generate and manipulate reference images interactively and locally synthesize diverse reference images onto artists' sketches. This process focuses on converting text-based design to visual expression, significantly enhancing communication clarity and specificity. We-toon generates reference images through GAN-based image synthesis and manipulation from a webtoon image corpus dataset. In this way, the quality and stability of the generated stylized images for webtoons are assured. The research with 24 webtoon authors implies that We-toon may outperform traditional methods in communication efficiency and satisfaction with revised images}\added{For example, We-toon is a system that bridges the contextual gap between webtoon writers and artists during sketch revisions. It allows writers to generate and manipulate reference images and synthesize them onto artists' sketches, enhancing communication clarity and specificity. Using GAN-based image synthesis from a webtoon image corpus, We-toon ensures high-quality stylized images. Research with 24 webtoon authors suggests that We-toon may outperform traditional methods in communication efficiency and satisfaction with revised images} \cite{ko2022we}. \removed{In contrast to this approach, in Sketchar, we use DALLE2, which generates images that are contextually coherent and responsive to prompts \cite{marcus2022very}. The reason for this is that complex elements like gameplay, world setting, character roles, and artistic styles can be prompted interactively using our approach. We also used GPT-3.5 to do structured extensions and additions to user-entered prompts, aiming to ensure a result that includes a polished game design document as well as reference images for illustrators.}\added{In Sketchar, we use DALLE instead of GAN, which can generate images in multiple styles that are contextually coherent and responsive to prompts \cite{marcus2022very}. This is because compared to webtoons, games are more different in complex elements like gameplay, world setting, character roles, and artistic styles, all of these can be prompted interactively using our approach during the game character design process. We also used GPT-3.5 to do structured extensions and additions to user-entered prompts, aiming to ensure a result that includes a polished game design document as well as reference images for illustrators.}

Like the We-toon study, we relied on a formative study to clarify the challenges and used baseline and treatment conditions to assess the application. However, in our study, we did not allow designers and illustrators to pass messages but rather created a task where they are to "hand off" the outcomes of the design to the next member of the team. Unlike in We-toon, the interactive prompts allow the participant to tailor the images to their liking, hence creating a reduced need for illustrators to preview the workflow. In essence, the GenAI process allows the designer to overtake a greater burden of image prototyping instead of relying only on image curation from a GAN model. We hypothesize that this more "self-sufficient" process allows designers to prototype the look more readily, giving them a greater opportunity to play with the look of the characters. The evidence for this is in both the qualitative findings (KF3) and the resulting higher CSI score in Collaboration compared to the baseline. However, more research is needed to probe the specific processes within the collaboration effort that would be specifically affected by text-to-image support.

A common challenge in game character design is the diverse styles in which game character design documents are written, often impeding collaboration and communication between game designers and illustrators (R4, R7, and R8). Sketchar, by translating game designers' input into detailed, standardized documentation with refined keywords (see Fig. 8), presents a possible solution to this issue. As one participant (P3) noted, \textit{"the well-organized design document can help me express my ideas to the game illustrators from the same group,"} highlighting Sketchar's potential to bridge the communication divide between game designers and illustrators.

Additionally, Sketchar's rapid generation of reference images offers designers a tool for effectively conveying their ideas visually. In the game industry, the designer always plays a significant role in approving the final outcomes of game character design and concept art before progressing to subsequent stages. Our study reveals that game designers are willing to use results produced by Sketchar as foundational blueprints for further refinement, and these blueprints are thought to be more detailed than if they draw them themselves (P2, P3, P5, P11). The generated game design document section is thought to be \textit{"more structured compared to the direct use of the Current generative AI tool"}(P4). At least some of the images produced were also evaluated to be valid for professional use by expert illustrators. Together, this potentially allows illustrators to more swiftly and accurately grasp the intended design concepts that were put forth by designers without illustration expertise. \added{Simultaneously, such structured generation would also aid game designers in organizing design documentation for subsequent development.
}

\added{\subsection{GenAI and the Shifts in the Landscape of Game Development}

The use of generative AI in creative fields has indeed introduced convenience somehow \cite{prediction,prediction1}, but it has also brought certain challenges and concerns. A study in 2023 \cite{boucher2024resistance} found significant skepticism among game developers about generative AI, with concerns about workflow integration, ownership, copyright, and ethical use. Programmers generally welcome AI tools, viewing code as functional. In contrast, artists are cautious, fearing copyright issues and loss of creative control, and worry that AI might replace human creativity. We found a similar contrast in our study. For generative AI tools to assist game design, many game artists tend to have a more negative attitude and concerns than game designers. As an artist (R6) stated: \textit{"I spent many years mastering the skills of character design, but AI can quickly learn these skills by simply studying existing data, including the data of characters I have designed."}

Our study suggested that AI tools support creative collaboration and communication in game development.  However, they cannot replace human input due to issues like lighting, character pose, and location accuracy.  Even if AI-generated content reaches a high quality one day, experienced artists still need to select suitable outputs. Only trained and experienced artists possess the refined aesthetic judgment necessary to choose the most appropriate outputs for specific game settings. We suggest that generative AI like Sketchar could change roles in game development. Artists may no longer need to design entirely by hand from scratch but can instead select and modify the best images designers generated by AI. We also propose more discussions between game companies and indie creators about data training and establishing data collection rules, because many artists are concerned about their data being used by large companies to train models without their knowledge \cite{stark2019work}.

Additionally, there are widespread concerns in the game development community about biases and stereotypes in AI-generated images \cite{AlgorithmWatch2024,Bloomberg2023}. Our study also revealed the stereotype effect due to DALLE-2 when designing characters from indigenous communities. For example, when drawing an image of a brave man from the Maori community in New Zealand, it ended up showing a white male with indigenous-style tattoos (\ref{fig:Maori}). When drawing a successful Chinese businessperson, the results are always a young Chinese male dressed in a suit. Biased training data may cause these stereotypes and biases. In such cases, using generative AI may exacerbate existing challenges related to the representation and cultural appropriation of minority groups and Indigenous cultures in the game industry. In the future, we plan to integrate professional cultural counselors into the Sketchar workflow. Suppose the system detects that users are trying to design game characters representing Indigenous people or minority groups. In that case, online crowd-sourced cultural counselors will review the generated content before being presented to users. When cultural counselors from the target community are unavailable, Sketchar will notify users that the generated content may contain biases and stereotypes and recommend consulting a professional cultural advisor before using the results. Besides, we proposed to conduct further discussions and collaborate with industry professionals to address and resolve current biases and stereotypes in AI models.}

\added{\begin{figure} 
    \centering 
    \includegraphics[width =0.50\textwidth]{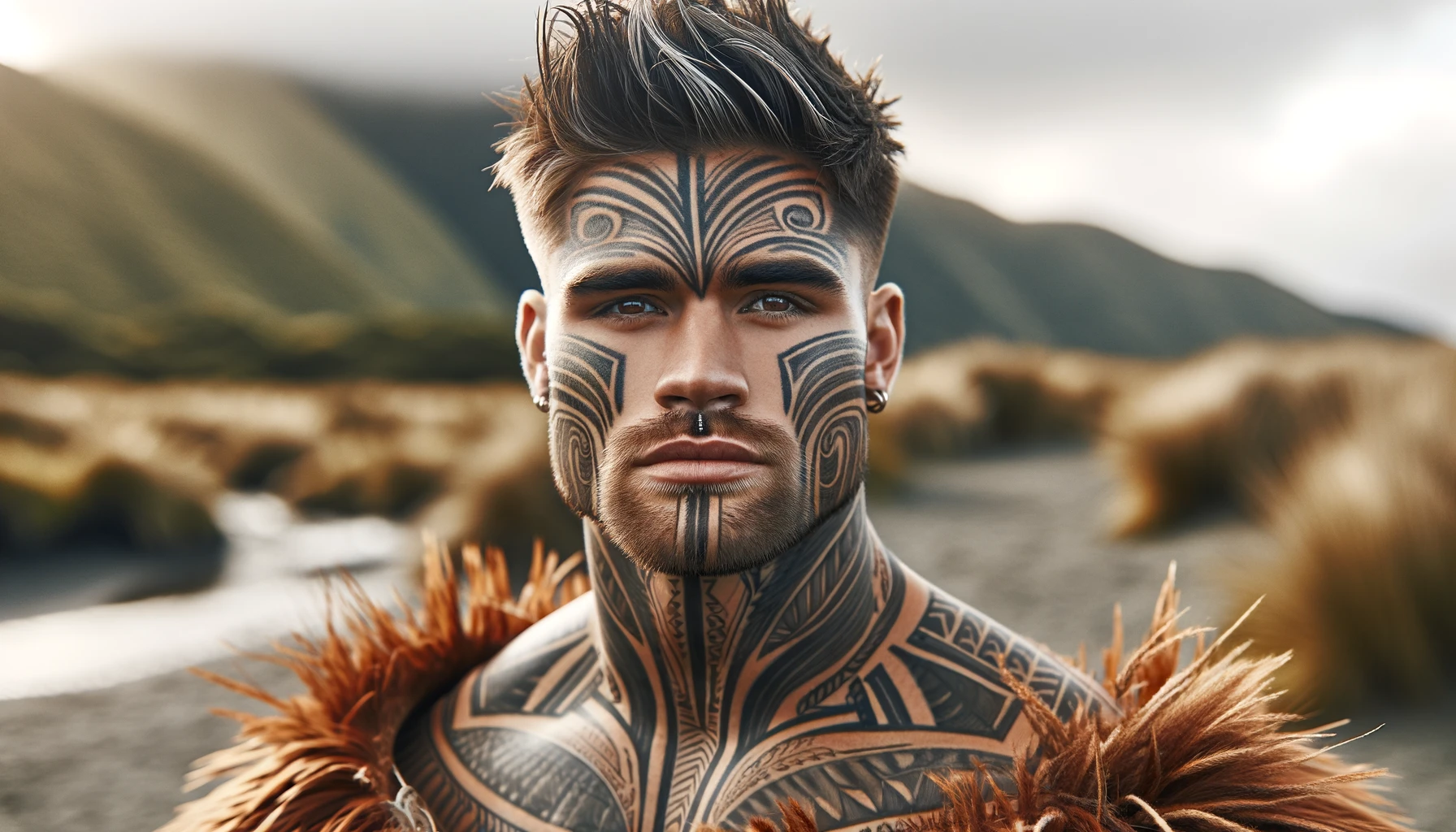} 
    \vspace{-0.24cm}
    \caption{Generated by DALLE, Prompt: \textit{A brave male from the Maori community in New Zealand},
} 
    \label{fig:Maori} 
    \vspace{-0.5cm}
\end{figure}}

\subsection{Limitation and Future Work}

One of the limitations of the current work is that some designers find it challenging to generate high-quality images using Sketchar because the platform sacrifices the intricacy of reference images for faster generation speed. For instance, illustrator P9 had to use 5 iterations to achieve an outcome with distinct facial tattoos during the actual image generation process, as illustrated in Figure \ref{results}(d). We think that switching from DALL-E 2 to a more controllable generative model, like Stable Diffusion or DALL-E 3, could improve the visual quality~\cite{borji2023generated,fernandez2022technology}. However, this might make the generation time longer, which might make the app less interesting. 

\added{Another limitation of this study is that Sketchar has only been tested with participants from mainland China. China has the world's largest gaming market, but government regulations and stigmatization have sometimes hindered local game development. As a result, Chinese developers have grown up in a unique environment and likely have different perceptions and understandings of game design. For example, in China, game designers have greater decision-making power and lead the game design process. Therefore, if we tested this system with Western participants, they may have a different workflow. Artists might engage more in the early stages of game character design,  and this would change the way that designers would interact with Sketchar. Additionally, during the study, ChatGPT and DALL-E were not accessible in China \cite{forbes2023chatgpt}, resulting in most participants lacking experience with generative AI. In contrast, previous studies involving Western participants \cite{boucher2024resistance,li2024user,inie2023designing} found that most had used generative AI, making them more aware of its limitations and risks. Consequently, involving Western participants in testing Sketchar could reveal more potential issues.}

Furthermore, this study only simulated collaboration during the game character design process. Despite sharing features for facilitating collaborative workflows between designers and illustrators, the tool currently lacks the capability for simultaneous content creation within the same workspace. It is worth noting that in our testing phase, single individuals worked with Sketchar, meaning that the collaboration had to be imagined to be taking place. For example, we told the designers working on Sketchar to imagine handing the prototyped images to an illustrator to refine and create actual game images. This does not capture the actual collaboration that would consist of iterative feedback and refinement in the actual game design collaboration process between designers and illustrators. Due to the lack of in-the-moment collaboration between these stakeholders, in the interviews, designer P4 expressed a desire for more collaborative interactions, stating, \textit{ "It would be great if we could work (physically) together with illustrators."} To address this issue, future user studies would involve recruiting game designers and experienced illustrators to work together on a task, as found in the We-toon study, which recruited webtoon writers to use the tool and artists to do evaluations \cite{ko2022we}. Following the completion of game character design documentation, including reference images by game designers with the help of Sketchar, illustrators would be enlisted to conduct evaluations grounded in their professional expertise instead of evaluating the images post hoc in the current study. Alternatively, we may look at the choices that designers made in deciding on the image to represent their characters to see what properties of the images make them acceptable, providing us with an idea of what aspects of the illustration prototypes worked and what did not. \added{Besides, in this study, we compared Sketchar only with participants' original workflows, without direct comparisons to generative AI tools like ChatGPT and DALL-E. This decision was due to the difficulty of accessing these tools in China during the study and preliminary interviews indicating that most participants lacked experience with generative AI. Additionally, our system is designed to synchronize visual content and text within the game design process in this study, we do not aim to compare them separately. Future research would include direct comparisons between Sketchar and ChatGPT plus DALL-E.}

In addition, the CSI Collaboration subscale is generally used to evaluate synchronous collaboration with another person, whereas Sketchar is designed to facilitate asynchronous collaboration between designers and illustrators, so the perception that the tool supports collaboration may be speculative. However, in real work scenarios, it is the designer who evaluates the final result of the character, and since our participants included designers with art skills, some of whom had experience as illustrators, their high evaluations of collaboration could reflect the effectiveness of Sketchar to some extent. 



\added{At last, in the future, we aim to explore the educational impacts of Sketchar. Specifically, we want to investigate whether the structured outputs and contextually appropriate reference images can improve designers' document organization and visual expression skills, helping them enhance their abilities by learning from AI-generated results. In this way, humans can in turn get enhanced by learning from AI generation.}

\section{Conclusion}\label{sec:Conclusion}
This study elucidates the collaborative process involved in game character creation and highlights the difficulties encountered in the communication between designers and artists. In response to this challenge, we developed Sketchar, a tool that utilizes GenAI to assist in the creation of game characters and facilitate the way designers may communicate character design ideas to illustrators. Following that, we carried out an assessment of Sketchar's efficacy in real-world game character creation situations with game designers.

Our study highlights the potential of GenAI to facilitate multidisciplinary cooperation by empowering designers to prototype utilizing creative drawing techniques that may exceed their current capabilities. This exemplifies how GenAI empowers individuals without specialized knowledge to prototype using approaches outside their domain of expertise.



\bibliographystyle{ACM-Reference-Format}
\bibliography{references}

\newpage
\newpage
\newpage
\appendix

\label{sec:Appendix}

\section{Appendix: The outline of the in-depth interview with character designers}

\subsection{Workflow}

\textbf{The background of the interviewers:}

How many years have you worked as a character designer?

Can you briefly describe your previous experiences as a character designer?

\textbf{The workflow of character designers:}

What is the first step of your workflow? How do you work with your original idea?

Please briefly describe your whole workflow.

How do you iterate the character you design?

\textbf{The collaboration and the tool:}

Have you worked with others on the same team as a character designer?

What kind of tools do you use to visualize the character you design?

\subsection{Character}

\textbf{The main consideration in character design:}

When brainstorming characters, what specific aspects do you primarily focus on, such as appearance and personality?

\textbf{The relationship of multiple characters:}

When it comes to designing multiple characters, how do you organize their relationships?

\textbf{The scene in the plots:}

When writing character storylines, do you consider factors such as geographical location, environment, weather, and time? If so, how do you envision such scenes?

\subsection{Requirements}

What aspects of the current tools you are using are you dissatisfied with? If there could be improvements, what additional features would you like to see?

Do you use AI software such as ChatGPT and Midjourney in your work? What aspects of their usage do you find unsuitable or lacking?

\section{APPENDIX: CSI SCALE COMTENT}

\begin{figure}[htbp]
 \vspace{-0.4cm}
  \centering
    \includegraphics[width=.55\linewidth]{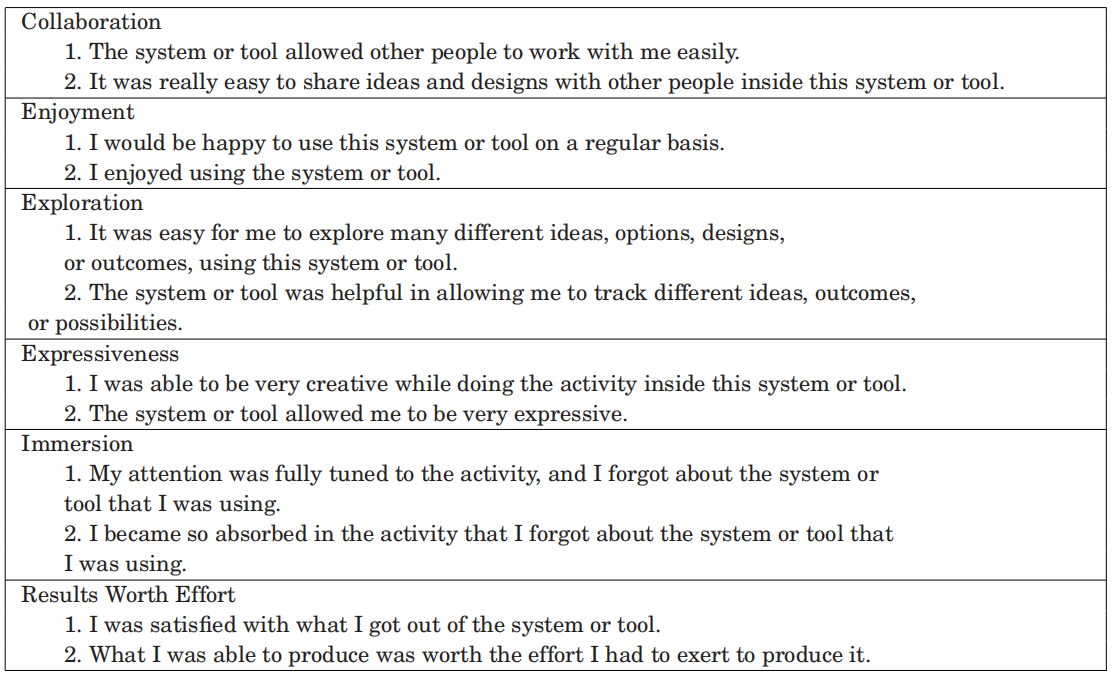}
  \vspace{-0.3cm}
  \caption{Six dimensions and specific descriptions of CSI. The answer for each item is "High Disagree" (0) to "High Agree" (10).}
 \label{csicontent}
  \vspace{-0.36cm}
\end{figure}

\end{document}